\documentclass[twoside,12pt]{article}
\usepackage{graphicx}

\def\simgt{\ {\raise-.5ex\hbox{$\buildrel>\over\sim$}}\ }
\def\simlt{\ {\raise-.5ex\hbox{$\buildrel<\over\sim$}}\ }

\begin{document}

\begin{huge}
\begin{center}
{\bf{Asteroseismology}}\\
\vspace{7mm}
\end{center}
\end{huge}

\thispagestyle{empty}
%\begin{large}
\begin{center}
Gerald Handler\\
Copernicus Astronomical Center, Bartycka 18, 00-716 Warsaw, Poland\\
Email: gerald@camk.edu.pl\\
\end{center}
%\end{large}

\section*{Abstract}

Asteroseismology is the determination of the interior structures of 
stars by using their oscillations as seismic waves. Simple explanations 
of the astrophysical background and some basic theoretical 
considerations needed in this rapidly evolving field are followed by 
introductions to the most important concepts and methods on the basis of 
example. Previous and potential applications of asteroseismology are 
reviewed and future trends are attempted to be foreseen.

%\subsection*{Index terms}

%Stellar structure, stellar evolution, pulsations, oscillations, variable stars, convection, normal modes, spherical harmonics, convective overshooting, stellar rotation, HR Diagram, photometry, spectroscopy, stellar models, binary stars, stellar clusters, radial velocities, frequency analysis, space satellites, mode identification.

%\subsection*{Key words}

%convection, hydrodynamics, line: profiles, magnetic fields, nuclear reactions, methods: data analysis, space vehicles, techniques: photometric, techniques: radial velocities, techniques: spectroscopic, stars: atmospheres, binaries: eclipsing, stars: chemically peculiar, stars: early-type, stars: evolution, stars: fundamental parameters, Hertzsprung-Russell diagram, stars: horizontal-branch, stars: interiors, stars: magnetic fields, stars: oscillations (including pulsations), stars: pre–main-sequence, stars: rotation, Cepheids, Delta Scuti stars, variables: other, white dwarfs

\section*{Introduction: variable and pulsating stars}

Nearly all the physical processes that determine the structure and
evolution of stars occur in their (deep) interiors. The production of
nuclear energy that powers stars takes place in their cores for most of
their lifetime. The effects of the physical processes that modify the
simplest models of stellar evolution, such as mixing and diffusion, also
predominantly take place in the inside of stars.

The light that we receive from the stars is the main information that 
astronomers can use to study the universe. However, the light of the 
stars is radiated away from their surfaces, carrying no memory of its 
origin in the deep interior. Therefore it would seem that there is no 
way that the analysis of starlight tells us about the physics going on 
in the unobservable stellar interiors.

However, there are stars that reveal more about themselves than others.
{\it Variable stars} are objects for which one can observe
time-dependent light output, on a time scale shorter than that of
evolutionary changes. There are two major groups of variable star,
extrinsic and intrinsic variables.

Extrinsic variables do not change their light output by themselves. For 
example, the light changes of eclipsing binary stars are caused by two 
stars passing in front of each other, so that light coming from one of 
them is periodically blocked. The individual components of eclipsing 
binary stars are not necessarily variable. By analysing the temporal 
light variations and orbital motion of eclipsing binaries, one can 
determine their fundamental properties, and by assuming that their 
components are otherwise normal stars, determine fundamental properties 
of all stars, most importantly their masses. In this way, stars and 
stellar systems can be understood better.

Intrinsic variables, on the other hand, change their light output
physically. Supernovae, which are stellar implosions/explosions, can
become brighter than their host galaxies because of the ejection of large
amounts of material. Even more revealing are stars that vary their sizes
and/or shapes: {\it pulsating variables}.

The first pulsating star was discovered more than 400 years ago. In 1596
David Fabricius remarked that the star $o$~Ceti (subsequently named
``Mira'', the wonderful) disappeared from the visible sky. About 40 years
later, it was realized that it did so every $\sim$~11 months; the first
periodic variable star was known (although we know today that in this
case, the term ``periodic'' is not correct in a strict sense).

In 1784 John Goodricke discovered the variability of $\delta$~Cephei, and
in 1914 enough evidence had been collected that Harlow Shapley was able to
demonstrate that the variations of $\delta$~Cephei and related stars (also
simply called ``Cepheids'') was due to radial pulsation. Also in the teens
of the previous century, Henrietta Leavitt pointed out that the Cepheids
in the Small Magellanic Clouds follow a period-luminosity relation, still
one of the fundamental methods to determine distances in the visible
universe - and one of the major astrophysical applications of pulsating
stars.

With the ever increasing precision in photometric and radial velocity
measurements, a large number of groups of pulsating star is nowadays
known. Figure 1 shows theoretical (in the sense that the logarithm of the
effective temperature is plotted versus the logarithm of the stellar
luminosity) HR Diagrams containing the regions in which pulsating stars
were known some 40 years ago and today. Table 1 gives a rough overview of 
the classes of pulsator contained in Fig.\ 1.

\begin{figure}[!htb]
\centering
\includegraphics[clip,angle=0,width=14.5cm]{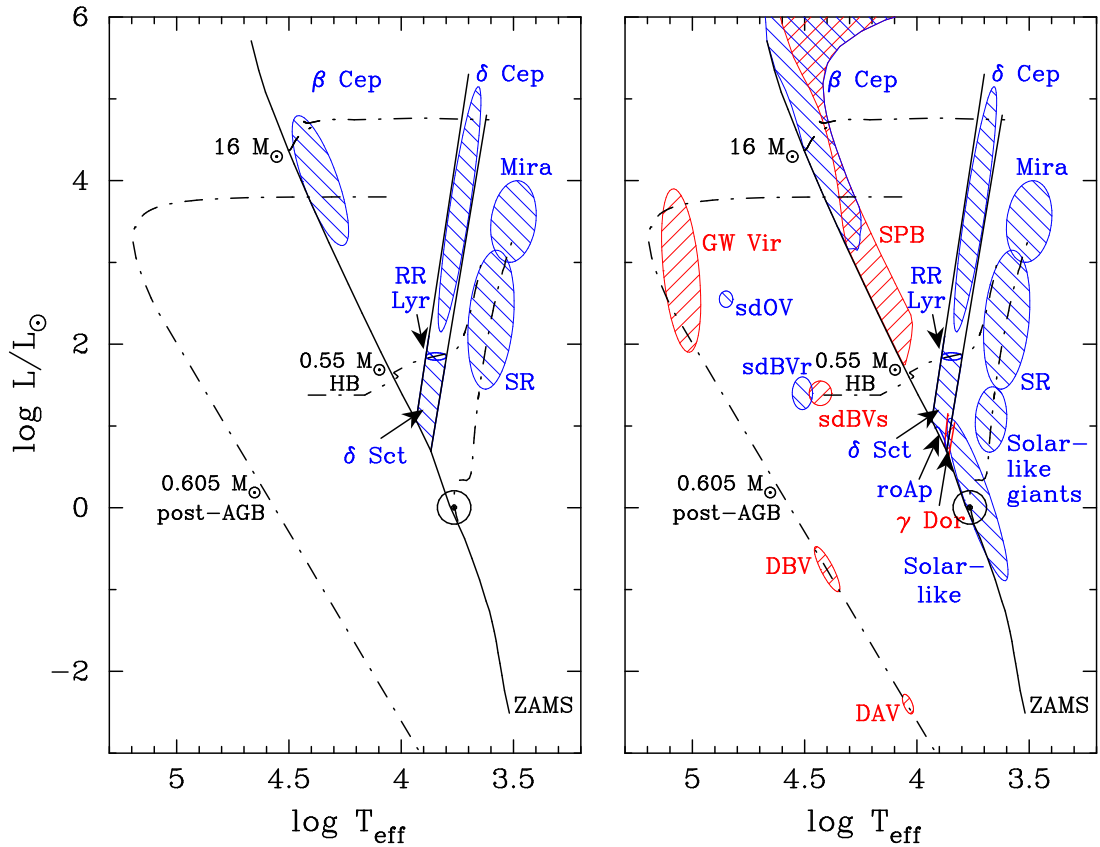}
\caption{Theoretical HR diagram schematically showing locations of 
selected confirmed types of pulsating star. Parts of model evolutionary 
tracks for main sequence, horizontal branch and post-AGB stars are shown 
as dashed-dotted lines for orientation. Left: known classes of pulsating 
star about 40 years ago. Right: classes of pulsator known to date. The 
names of the different groups are comprehensively listed in Table 1. Note 
the enormous increase in the number of classes of pulsating star and the 
improved knowledge about their loci in the recent past.}
\end{figure}

\begin{table}
\caption[]{Selected classes of pulsating star}
\begin{center}
\begin{tabular}{lcl}
\hline
Name & Approx.\ Periods & Discovery/Definition\\
\hline
Mira variables & 100 - 1000 d & Fabricius (1596)\\
Semiregular (SR) variables & 20 - 2000 d & Herschel (1782)\\
$\delta$ Cephei stars & 1 - 100 d & 1784, Pigott, Goodricke (1786)\\
RR Lyrae stars & 0.3 - 3 d & Fleming (1899)\\
$\delta$ Scuti stars & 0.3 - 6 h & Campbell \& Wright (1900)\\
$\beta$ Cephei stars & 2 - 7 h & Frost (1902) \\
ZZ Ceti stars (DAV) & 2 - 20 min & 1964, Landolt (1968)\\
GW Virginis stars (DOV) & 5 - 25 min & McGraw et al. (1979) \\
Rapidly oscillating Ap (roAp) stars & 5 - 25 min & 1978, Kurtz (1982)\\
V777 Herculis stars (DBV) & 5 - 20 min & Winget et al. (1982) \\
Slowly Pulsating B (SPB) stars & 0.5 - 3 d & Waelkens \& Rufener (1985)\\
Solar-like oscillators & 3 - 15 min & Kjeldsen et al. (1995)\\
V361 Hydrae stars (sdBVr) & 2 - 10 min & 1994, Kilkenny et al. (1997)\\
$\gamma$ Doradus stars & 0.3 - 1.5 d & 1995, Kaye et al. (1999)\\
Solar-like giant oscillators & 1 - 18 hr & Frandsen et al. (2002)\\
V1093 Herculis stars (sdBVs) & 1 - 2 hr & Green et al. (2003)\\
Pulsating subdwarf O star (sdOV) & 1 - 2 min & Woudt et al. (2006)\\
\hline
\end{tabular}
\end{center}
\end{table}

The different types of pulsator have historically been classified on a 
phenomenological basis. The separation between those types has usually 
later turned out to have a physical reason. The individual classes are 
different in terms of types of excited pulsation mode (or, less 
physical, pulsation period), mass and evolutionary state, hence 
temperature and luminosity. The names of these classes are assigned 
either after the name of a prototypical star or give some description of 
the type of variability and star.

It must be pointed out that the present overview does by far not contain 
all types and subgroups of pulsating star that have been suggested. The 
Cepheids are subdivided according to population and evolutionary state, 
into $\delta$~Cephei, W Vir, RV Tau and BL Her stars. Jeffery (2008) 
proposed a number of types of evolved variable, there are the Luminous 
Blue Variables, and there may be new classes of white dwarf pulsator, 
oscillating red and brown dwarfs, etc. Furthermore, some of the 
instability domains of different pulsators overlap and indeed, some 
objects called ``hybrid'' pulsators that show oscillations of two 
distinct types, have been discovered. Also, the instability boundaries 
of some of these variables may need to be (considerably) extended and/or 
revised in the near future. For instance, there may be supergiant SPB 
stars, and solar-like oscillations are expected in all stars having a 
significant surface convection zone.

Whereas the classification of and distinction between the different 
classes of pulsating star, that are historically grown and modified 
designations, can in some cases be called arbitrary today, one 
recognizes an important fact: pulsating stars populate almost the entire 
HR diagram, and this means that they can be used to learn something 
about the physics of most stars.

\section*{Astrophysical background}

\subsection*{Driving mechanisms}

What can make a star oscillate? After all, stars are in hydrostatic 
equilibrium: the gravitational pull on the mass elements of normal stars 
is balanced by gas pressure. If something would hit a star, the inwards 
moving regions will be heated, and the increased heat loss damps the 
motion. Consequently, self-excited pulsations require a driving 
mechanism that overcomes this damping and results in a periodic 
oscillation.

Four major driving mechanisms have been proposed. The {\it $\epsilon$ 
mechanism} (Rosseland \& Randers 1938) assumes a variation in the 
stellar nuclear reaction rate: if a nuclear burning region is 
compressed, the temperature rises and more energy is produced. This 
gives causes expansion, the pressure drops, and so does the energy 
generation: the motion is reversed and oscillations develop. The 
$\epsilon$ mechanism (where $\epsilon$ is the usual designator for the 
nuclear reaction rate in formulae), that operates similar to a Diesel 
engine, has been proposed for several different types of pulsating star, 
such as our Sun and pulsating white dwarfs, but observational proof for 
oscillations driven by it is still lacking.

Considerably more successful in explaining stellar oscillations is the 
{\it $\kappa-\gamma$ mechanism} (Baker \& Kippenhahn 1962 and references 
therein). In layers where the opacity $\kappa$ increases and/or the 
third adiabatic exponent $\Gamma_3$ decreases with increasing 
temperature, flux coming from inner layers can be temporally stored. 
Such layers in the stellar interior are generally associated with 
regions where (partial) ionization of certain chemical elements occurs.

The energy accumulated in this layer during compression is additionally 
released when the layer tries to reach its equilibrium state by 
expanding. Therefore, the star can expand beyond its equilibrium radius. 
When the material recedes, energy is again stored in the stellar 
interior, and the whole cycle repeats: a periodic stellar oscillation 
evolves. This mechanism is also called the {\it Eddington Valve}, and it 
explains the variability of most of the known classes of pulsating star. 

The classical pulsators in the instability strip, ranging from the 
$\delta$~Cephei stars to the RR Lyrae stars and the $\delta$ Scuti stars 
draw their pulsation power from the He{\sc ii} ionization zone, whereas 
the oscillations of the roAp stars are believed to be excited in the 
H{\sc i} and He{\sc i} ionization zones, those of the Mira variables in 
the H{\sc i}ionization zone, and those of the $\beta$~Cephei and SPB 
stars are triggered in the ionization zone of the iron-group elements.

A very similar mechanism, in the sense that it is also due to a region in
the star behaving like a valve, is convective blocking (or {\it convective
driving}). In this scheme (e.g., Brickhill 1991), the base of a convection
zone blocks the flux from the interior for some time, releasing the energy
stored during compression in the subsequent expansion phase. The
pulsations of white dwarf stars of spectral types DA and DB as well as
$\gamma$~Doradus stars are thought to be excited (at least partly) via
this mechanism, that may also be of importance in Cepheids and Mira 
stars.

Finally, the pulsations of the Sun and solar-like stars, that are 
intrinsically stable and therefore not called self-excited, are {\it 
stochastically excited} due to turbulence in their surface convection 
zones. The vigorous convective motion in the outer surface layers 
generates acoustic noise in a broad frequency range, which excites 
solar-like oscillation modes. Due to the large number of convective 
elements on the surface, the excitation is of random nature, and the 
amplitudes of the oscillations are temporally highly variable.

Given the physical nature of these driving mechanisms, the existence of 
the different instability domains in the HR diagram (cf.\ Fig.\ 1) 
easily follows. A star must fulfil certain physical conditions that it 
can pulsate, as the driving mechanism must be located in a specific part 
of the star to give rise to observable oscillations. More physically 
speaking, in the case of self-excited pulsations the driving region must 
be located in a region where the thermal, and/or convective time scale 
closely corresponds the dynamical (pulsational) time scale.

The consequence of the previous requirement is a constraint on the 
interior structure of a pulsating variable: if the instability region of 
some class of pulsating variable is accurately known, models of stars 
incorporating its excitation mechanism must be able to reproduce it. In 
this way, details of the input physics describing the interior 
structures of stars can be modified to reflect the observations.

However, this is not the only method available to study stellar structure
and evolution from stellar pulsation. Models also need to explain the
oscillation properties of individual stars. We are fortunate to be in the
presence of stars having very complex pulsation patterns, multiperiodic
radial and nonradial oscillators. The research field that determines the
internal constitution of stars from their pulsations is called
{\it asteroseismology}.

It is now due to make clear that the star whose interior structure is 
best known is the star closest to us, the Sun. Its surface can be 
resolved in two dimensions and millions of pulsation modes can be used 
for seismic analyses. The related research field is {\it 
helioseismology}, and it has been extensively reviewed elsewhere (e.g., 
Christensen-Dalsgaard 2002, Gizon et al.\ 2010). The present article 
will not touch upon helioseismology.

\subsection*{Asteroseismology}

The basic idea of asteroseismology (see Gough 1996 for a discussion of 
why ``astero") is analogous to the determination of the Earth's inner 
structure using earthquakes: these generate seismic waves that travel 
through rock and other interior structures of our planet, which can then 
be sounded. Today, the Earth's interior has been completely mapped down 
to scales of a few hundred kilometres.

The strongest earthquakes can even cause {\it normal mode oscillations} 
(e.g., see Montagner \& Roult 2008), that is, the whole planet vibrates 
with its natural frequencies (also called {\it eigenfrequencies} in 
theoretical analyses). These normal modes are most valuable in 
determining the deep interior structure of the Earth. Asteroseismology 
does just the same: it uses the frequencies of the normal modes of 
pulsating stars (that may be seen as ``starquakes'') as seismic waves. 
The eigenfrequencies of stellar models, that are dependent on their 
physical parameters and interior structures, are then matched to the 
observed ones. If a model succeeds in reproducing them, it is called a 
{\it seismic model}.

The pulsation modes are waves in the stellar interior, just like the 
waves that musical instruments resonate in. Popular articles often 
compare the frequency pattern of stellar oscillations to the sounds of 
musical instruments. The superposition of the normal mode frequencies in 
which a star oscillates can therefore be seen as its sound, although it 
generally does not involve frequencies the human ear is susceptible to.

There is a large variety of normal modes that stars can pulsate in.
The simplest are {\it radial modes}: the star periodically expands and 
shrinks, and its spherical symmetry is preserved. The mathematical 
description of the displacement due to the oscillations results in 
differential equations of the Sturm-Liouville type that yields
discrete eigensolutions: the radial mode frequencies of the given model.

Pulsation in {\it nonradial modes} causes deviations from spherical 
symmetry: the star changes its shape. Mathematically, this no longer 
results in an eigenvalue problem of the Sturm-Liouville type, and a 
large number of possible oscillation modes originates. The 
eigenfunctions are proportional to {\it spherical harmonics}:

\begin{equation}
Y_l^m(\theta,\phi)=N_l^m P_l^{|m|}(\cos\theta)e^{im\phi},
\end{equation}

where $\theta$ is the angle from the polar axis (colatitude), $\phi$ is 
the longitude, $P_l^{|m|}$ is the associated Legendre polynomial, 
$N_l^m$ is a normalization constant and $l$ and $m$ are the spherical 
degree and azimuthal order of the oscillation. With

\begin{equation}
P_l^m(x)=(-1)^m(1-x^2)^{m/2}\frac{d^m}{dx^m}P_l(x)
\end{equation}

we obtain

\begin{equation}
P_0^0(\cos\theta)=1
\end{equation}
\begin{equation}
P_1^0(\cos\theta)=\cos\theta \hspace{24mm} 
P_1^1(\cos\theta)=-\sin\theta
\end{equation}
\begin{equation}
P_2^0(\cos\theta)=\frac{1}{2}(3\cos^2\theta-1) \hspace{12mm} 
P_2^1(\cos\theta)=-3\cos\theta\sin\theta \hspace{12mm} 
P_2^2(\cos\theta)=3\sin^2\theta
\end{equation}

\begin {center}
etc. etc.
\end {center}

What does this mean in practice? Nonradial pulsation modes generate
distortions on the stellar surface described by these spherical harmonics.
The oscillations separate the stellar surface into expanding and receding
as well as heating or cooling areas. A graphical example of these is shown
in Fig.\ 2.

\begin{figure}[!htb]
\centering
\includegraphics[angle=0,width=15.0cm]{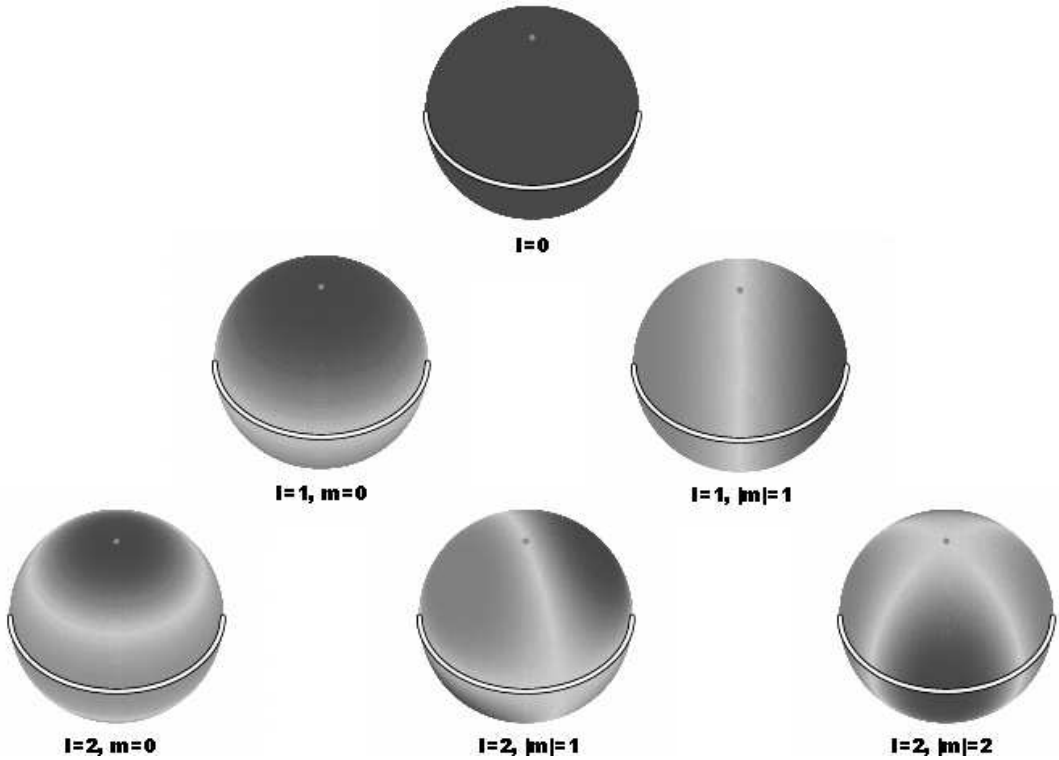}
\caption{Schematic description of the surface distortions produced by
pulsation modes with spherical degrees $0 \leq l \leq 2$, arranged in the
same way as the previous expressions for the associated Legendre
polynomials. Whilst the outward moving areas of the star are coloured in
dark grey, the light grey areas move inward, and vice versa. The pole and
equator of the star are indicated. Adapted from Telting \& Schrijvers
(1997).}
\end{figure}

Between the expanding and receding surface areas, no motion takes place. 
The lines along which this is the case are called the {\it node lines}; 
the nulls of the previously specified associated Legendre polynomials
specify the locations of these node lines in polar direction. The total
number of node lines on the stellar surface is the {\it spherical degree}
$l$, which must in all cases be larger than or equal to zero. The number
of node lines that are intersected when travelling around the stellar
equator is the {\it azimuthal order} $m$. $l$ and $m$ are the quantities
that appear in the expressions for the eigenfunctions and for the
associated Legendre polynomial given above, and they are used for the
classification of the pulsation modes.

Pulsation modes with $m \not= 0$ are travelling waves (as can also be 
seen in the defining equation for spherical harmonics), and as they can 
run either with or against the rotation of the star, m can lie in the 
interval $[-l,l]$. Modes with $m=0$ are called {\it axisymmetric}, modes 
with $|m|=l$ are named {\it sectoral}, and all other modes are referred to 
as {\it tesseral} modes.

The third quantity needed to describe pulsation modes is the {\it radial 
overtone} $k$ (sometimes also denoted $n$ in the literature), which is 
the number of nodes in the stellar interior. A mode that has no node in 
the interior is called a {\it fundamental mode}. A mode with one 
interior node is called the {\it first overtone}, modes with two 
interior nodes are the {\it second overtone}, etc. An accurate account 
of mode classification from the theoretical point of view is given by 
Deubner \& Gough (1984).

Historically, observationally, and inconsistently with the theoretical
definition, radial overtone modes have also been called the first and
second harmonics, respectively, and have been abbreviated with F for the
fundamental, as well as 1H (or 1O), 2H (or 2O) etc. for the overtones.
Radial pulsations can be seen as modes with $\ell=0$ (remember that
$P_0^0(\cos\theta)=1$); all other modes are nonradial oscillations. Modes
with $l=1$ are also called {\it dipole modes}; $l=2$ modes are {\it
quadrupole modes}.

There are two major restoring forces for stellar oscillations that 
attempt to bring the star back in its equilibrium configuration: 
pressure and buoyancy (gravity). For radial motion, the gravitational 
force in a star increases during compression, so it would actually 
accelerate, and not restore, the oscillation. Therefore, pressure must 
be the restoring force. On the other hand, for a predominantly 
transverse motion, gravity restores the motion through buoyancy, similar 
to what can be observed when throwing a stone into a pond. Therefore, 
aside from their identification with the pulsational quantum numbers $k, 
l$ and $m$, nonradial pulsation modes are also classified into p {\it 
pressure (p) modes} and {\it gravity (g) modes}. These two sets of modes 
thus differ by the main direction of their motion (radial/transverse), 
and their frequencies. Pulsation modes with periods longer than that of 
the radial fundamental mode are usually g modes, whereas p modes have 
periods equal or shorter than that; radial pulsations are always p 
modes.

The different modes are often labelled with their radial overtone 
number, e.g.\ a p$_3$ mode is a pressure mode with three radial nodes, 
and a g$_8$ mode is a gravity mode with eight radial nodes. Modes with 
no interior nodes are fundamental modes, or f modes. Note that the f 
mode for $l=1$ does not exist, as a dipole motion of the entire star 
would require a movement of the stellar centre of mass, which is 
physically impossible.

The propagation of pulsation modes in the stellar interior is governed 
by two frequencies. One of these is the {\it Lamb frequency} $L_l$, 
which is the inverse of the time needed to travel one horizontal 
wavelength at local sound speed. The other frequency describes at what 
rate a bubble of gas oscillates vertically around its equilibrium 
position at any given position inside a star; it is called the {\it 
Brunt-Vais{\"a}l{\"a}} frequency $N$. These two quantities are defined 
as:

\begin{equation}
L_l^2=\frac{l(l+1)c^2}{r^2} \hspace{30mm}
N^2=g\left(\frac{1}{p_0\gamma_1}\frac{dp_0}{dr}-\frac{1}{\rho_0}\frac{d\rho_0}{dr}\right),
\end{equation}

where $l$ is the spherical degree, $c$ is the local velocity of sound, 
$r$ is the radius, $g$ is the local gravitational acceleration, $p_0$ 
and $\rho_0$ are local pressure and density in the unperturbed state, 
respectively, and $\gamma_1=(\rho dp/pd\rho)_{\rm ad}$ is the first 
adiabatic exponent.

The Lamb and Brunt-Vais{\"a}l{\"a} frequencies have the following 
implications: an oscillation with a frequency higher than both 
experiences pressure as the main restoring force in the corresponding 
part of the star. On the other hand, a vibration with a frequency lower 
than both $L_l$ and $N$ is restored mostly by buoyancy. In other words, 
if we have a stellar oscillation with an angular frequency $\omega$, it 
is a p mode wherever $\omega > L_l, N$, and it is a g mode wherever 
$\omega < L_l, N$. In stellar interior regions where $\omega$ lies 
between the Lamb and Brunt-Vais{\"a}l{\"a} frequencies, the amplitude of 
the wave decreases exponentially with distance from the p and g mode 
propagation regions; such parts in the stellar interior are called 
evanescent regions. A {\it propagation diagram} aids the visualization 
of this discussion (Fig.\ 3).

\begin{figure}[!htb]
\centering
\includegraphics[width=11.5cm,clip]{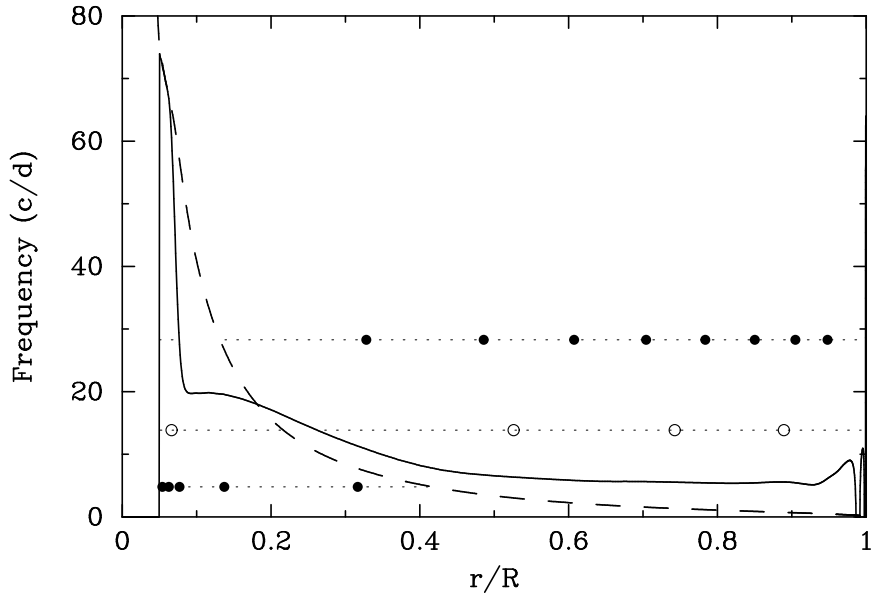}
\caption{A propagation diagram for an evolved 2 $M_{\odot}$ stellar model.  
The run of the Lamb (dashed line) and Brunt-Vais{\"a}l{\"a} (full line) 
frequencies with respect to fractional stellar radius is shown. Some
stellar pulsation modes are indicated with thin horizontal dashed lines,
the circles are interior nodes. The lowest frequency oscillation (lowest 
dashed line) shown is a $g_5$ mode; that with the highest frequency
is a $p_8$ mode. The oscillation with intermediate frequency is a mixed
mode. Data kindly supplied by Patrick Lenz.}
\end{figure}

Whereas the Lamb frequency decreases monotonically towards the model's 
surface, the Brunt-Vais{\"a}l{\"a} frequency shows a sharp peak near the 
model's centre and then rapidly drops to zero. This is because this 
model possesses a convective core, where $N^2<0$, and the spike in $N$ 
is due to a region of chemical inhomogeneity. Over a range of stellar 
models, the behaviour of $L_l$ in the interior is usually simple, 
whereas $N$ may show considerable changes with evolutionary state and 
mass. However, $N$ is always zero in the stellar centre.

In Fig.\ 3, the g mode is confined to the innermost parts of the star. 
It is trapped in the interior, and therefore unlikely to be observed on 
the surface. The p mode is concentrated near the stellar surface and may 
be observable. The intermediate frequency mode shows remarkable 
behaviour: it has three nodes in the outer regions of the star, but also 
one node in the g mode propagation region. This particular mode is 
capable of tunnelling through its narrow evanescent region; it is a g 
mode in the deep interior, but a p mode closer to the surface. Such 
modes are called {\it mixed modes}.

As mentioned before, stellar pulsation modes can be excited in certain 
parts of the stellar interior and they can propagate in some regions, 
whereas in other regions they are damped. The {\it work integral} is the 
energy gained by the pulsation mode averaged over one oscillation 
period. An evaluation of the work integral from the stellar centre to 
the surface is used to determine whether or not a given mode is globally 
excited in a stellar model. For excitation to occur, the exciting forces 
must overcome those of damping and the work integral $W$ will be 
positive.

The {\it growth rate} $\gamma=W/2\pi \omega I$, where $\omega$ is the 
pulsation frequency and $I$ is the mode inertia, parametrizes the 
increase of oscillation energy during a pulsation cycle, and also 
indicates how rapidly the amplitude of a given mode increases. The 
normalized growth rate (or stability parameter) $\eta$, which is the 
ratio of the radius-integrated work available for excitation to the 
radius-integrated total work, is used to evaluate which pulsation modes 
are excited in the given model. If $\eta>0$, a mode is driven and may 
reach observable amplitude; if $\eta=1$, a mode is driven in the entire 
stellar model, and if $\eta=-1$, a mode is damped everywhere in the 
model. The most widely used application of this stability parameter is 
the comparison of the excited modal frequency ranges as determined from 
observations and those predicted by theory.

Figure 4 shows how the frequency domains predicted to be excited change 
with the evolution of a 1.8\,$M_{\odot}$ main sequence model. On the 
Zero-Age Main Sequence (ZAMS), only pure p modes of high radial 
overtone, and corresponding high frequencies, are excited in the model. 
Later on, some of these become stable again, but modes with lower 
overtone become excited. At the end of main sequence evolution, a large 
range of p modes, mixed modes and g modes is predicted to be excited, 
and the frequency spectrum becomes dense (and even denser as the model 
leaves the main sequence). The range of excited frequencies is an 
observable that, when compared with models, can give an estimate of the 
evolutionary state of the star.

\begin{figure}[!htb]
\centering
\includegraphics[clip,angle=0,width=14.0cm]{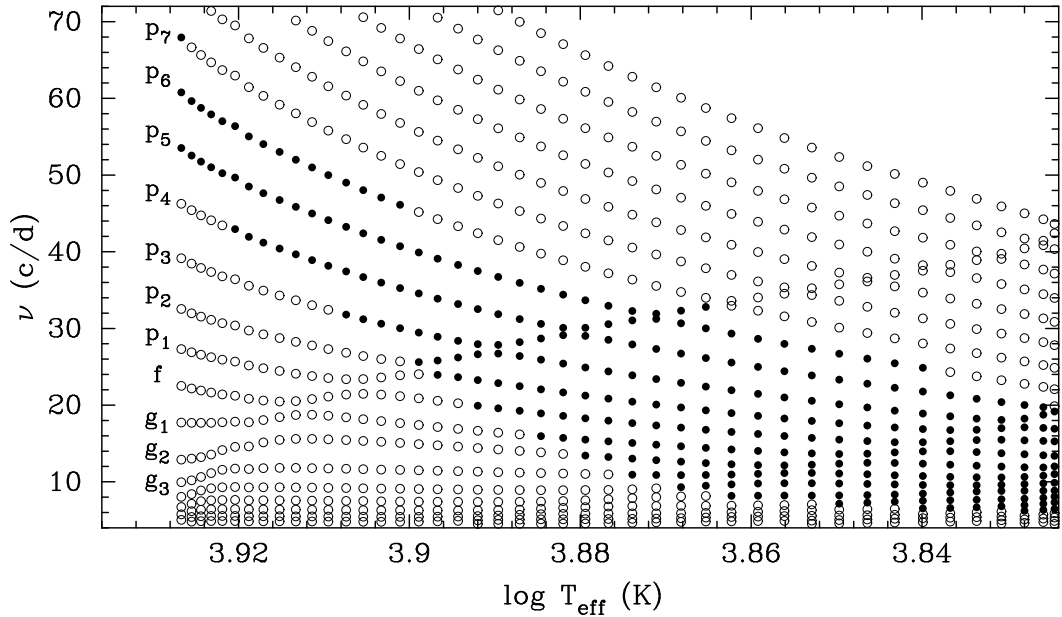}
\caption{Theoretically predicted $l=2$ oscillation spectra of a 
1.8\,$M_{\odot}$ main sequence model, evolving from hotter to cooler 
effective temperature. Pulsation modes excited in this model are shown 
with filled circles, stable modes with open circles. The types of
mode on the Zero-Age Main Sequence (ZAMS) are given. Note the g modes 
intruding into the p mode domain.}
\end{figure}

The reason for the evolutionary change in the pulsation frequencies is 
that as the model evolves, its convective core, in which the g modes are 
trapped at the ZAMS, shrinks and the g mode frequencies increase. At the 
same time the envelope expands due to the increased energy generation in 
the contracting nuclear burning core, causing a decrease of the p mode 
frequencies.

At some point, some p and g modes attain the same frequencies, and the 
modes begin to interact: they become mixed modes, with g mode character 
in the core and p mode character in the envelope (see also Fig.\ 3). The 
frequencies of these modes never reach exactly the same value; the modes 
just exchange physical properties. This effect is called {\it avoided 
crossing} or mode bumping (Aizenman, Smeyers \& Weigert 1977). 

Of the individual modes, particular astrophysical potential is carried 
by the mode that originates as $g_1$ on the ZAMS, as pointed out by 
Dziembowski \& Pamyatnykh (1991). This mode is mostly trapped in the 
shrinking convective core (for stars sufficiently massive to possess a 
convective core), because the rapid change in mean molecular weight at 
its edge causes a spike in the Brunt-Vais{\"a}l{\"a} frequency (cf.\ 
Fig. 3). The frequency of this mode is thus dependent on the size of the 
convective core. 

The most important parameter determining the evolution of main sequence 
stars is the amount of nuclear fuel available. Stellar core convection 
can mix material from the radiative layer on top of the core into it, 
providing more nuclear fuel. This mixing is often named {\it convective 
core overshooting} and parametrized in theoretical models. Overshooting 
also decreases the gradient in the mean molecular weight at the edge of 
the core, which means that the frequency of the mode that is $g_1$ on 
the ZAMS measures the convective core size, a most important quantity 
for astrophysics in general. To emphasize its sensitivity to the extent 
of stellar core convection, this mode has also been named $g_c$ mode.

In the following, only p and g modes, and mixed modes of these types will
be considered. Other types of mode, such as r modes (torsional
oscillations that may occur in rotating stars), g$^-$ modes (convectively
excited g modes in rotating stars), strange modes (showing up in
calculations of highly nonadiabatic environments) or gravitational-wave w
modes, will not be discussed as they have been of little practical
importance for the seismic sounding of stars at the time of this writing.

The frequencies of the p and g modes of pulsating stars depend strongly on
their structure. However, when high radial overtones are considered, some
simple relations between mode frequencies emerge. These are derived from
{\it asymptotic theory}. The classical reference on the subject is Tassoul 
(1980), the instructive is Gough (1986). In the high-overtone limit 
one finds

\begin{equation}
\omega_{n,l}\simeq\Delta\omega_0\left(n+\frac{l}{2}+\epsilon\right)-D_0l(l+1)
\end{equation}

for p modes, whereas for g modes

\begin{equation}
P_{n,l}\simeq\frac{P_0(n+\delta)}{\sqrt{l(l+1)}},
\end{equation}

where $\Delta\omega_0$ is the inverse sound travel time through the 
centre of the star, $D_0$ is a frequency separation dependent on the 
stellar evolutionary state, $P_0$ is the asymptotic period that is 
proportional to the integral of the Brunt-Vais{\"a}l{\"a} frequency 
throughout the star, $l$ and $n$ are the spherical degree and radial 
overtone, respectively, and $\epsilon$ and $\delta$ are stellar 
structure parameters, respectively.

These relations have important consequences: low-degree p modes of 
consecutive high radial overtones of the same spherical degree are 
equally spaced in frequency, whereas low-degree g modes of consecutive 
high radial overtones of the same spherical degree are equally spaced in 
period. Furthermore, if the parameter $D_0$ were zero, Eq.\ 7 indicates 
that the frequencies of high-overtone p modes of even and odd degrees 
would be the same, respectively, and odd-degree modes would have 
frequencies intermediate between those of even degree. In realistic 
stellar models, these relations hold approximately, but not exactly.

In a nonrotating star, the frequencies of modes with $m \neq 0$ are the 
same as those of the $m=0$ modes. However, as the $m \neq 0$ modes are 
travelling waves, their frequencies separate in the observer's frame 
when looking at a rotating star: the mode moving with rotation appears 
at higher frequency, the mode moving against rotation appears at lower 
frequency and the frequency difference to their nonrotating value is m 
times the rotation frequency (e.g., see Cox 1984). This effect is called 
{\it rotational frequency splitting}, and is one basic tool of 
asteroseismology: if such splittings are observed, the rotation 
frequency of the star can be determined.

Unfortunately, reality is not quite as simple as that. The Coriolis 
force acts on the travelling waves and modifies their frequencies. In 
addition, the $m \neq 0$ modes cause tidal bulges, on which centrifugal 
forces act. Therefore, the frequencies of these modes are often 
expressed as

\begin{equation}
\omega_{k,l,m}=\omega_{k,l,0} + m(1-C_{k,l})\Omega + 
m^2\frac{D_{k,l}\Omega^2}{\omega}
\end{equation}

in case of (moderately) slow stellar rotation (Dziembowski \& Goode 
1992). $\omega_{k,l,m}$ is the observed frequency of the mode $k,l,m$, 
$\Omega$ is the stellar rotation frequency, and $C_{k,l}$ and $D_{k,l}$ 
are constants that describe the effects of the Coriolis and centrifugal 
forces, respectively; they are also called Ledoux constants. These 
constants are usually determined from stellar model calculations. The 
rotational splitting constant $C_{k,l}$ also approaches asymptotic 
values for high radial overtones. For g modes, $C_{k,l}$ becomes 
$(l(l+1))^{-1}$, and for p modes $C_{k,l}=0$ in the asymptotic limit.

It should be made clear that the previous formula is only an 
approximation for the case that rotation can be treated as a 
perturbation to the equilibrium state: stellar rotation distorts the 
spherical shape of a star. As a consequence, the individual modes can no 
longer be described with single spherical harmonics. For instance, 
radial modes receive contamination of $l=2$ and other modes with even 
$l$. Vice versa, $l=2$ modes obtain some $l=0$, $l=4$, etc.\ 
contributions. In the previous formula this means that for rapid 
rotation higher order terms may be added, but to arrive at reliable 
results, two-dimensional numerical calculations are really required. 
{\it Rotational mode coupling} can affect the properties of oscillation 
modes with close frequencies (e.g., see (Daszy{\`n}ska-Daszkiewicz et 
al. 2002). Furthermore, the rotational distortion causes the stellar 
temperature to increase at the flattened poles and to decrease at the 
equatorial bulge. It follows that asteroseismology of rapidly rotating 
stars has an additional degree of complexity.

Even more complexity to the theoretical treatment of stellar pulsation 
is added by the presence of a magnetic field. A weak magnetic field 
would generate a second-order perturbation to the pulsation frequencies, 
just like the centrifugal force in case of rotation, but with opposite 
sign, i.e.\ the observed oscillation frequencies would increase with 
respect to the non-magnetic case (e.g., Jones et al.\ 1989).

In the presence of a strong magnetic field, the effect would be more 
severe: if the field was oblique to the rotation axis (just like the 
Earth's magnetic field!), the pulsation axis would align with the 
magnetic axis, and no longer with the rotation axis, as implicitly 
assumed so far. This means that an observer would see each pulsation 
mode at a varying angle over the stellar rotation period, creating 
amplitude and phase variations with exactly that period, and therefore 
separable from the effects of rotational m-mode splitting. A star 
oscillating in this way is called {\it oblique pulsator} (Kurtz 1982). 
We note for completeness that the most general case of rotational 
splitting is the {\it nonaligned pulsator} (Pesnell 1985), where 
$(2l+1)^2$ components of a given mode appear, with frequency separations 
proportional to the rotation period and to the angle between the 
rotation and pulsation axes.

To summarize, different stellar pulsation modes propagate in different 
interior regions, and their energy within those regions is not equally 
distributed. Each single pulsation mode has a different cavity, and its 
oscillation frequency is determined by the physical conditions in its 
cavity. This means that different modes are sensitive to the physical 
conditions in different parts of the stellar interior. Some modes teach 
us more about stellar envelopes, whereas other modes tell us about the 
deep interior. The more modes of different type are detected in a given 
star, the more complete our knowledge about its inner structure can 
become.

Some stars do us the favour to oscillate in many of these radial and/or 
nonradial modes simultaneously. Interior structure models of the stars 
can then be refined by measuring the oscillation frequencies of these 
stars, identifying them with their pulsational quantum numbers, and by 
reproducing all these with stellar models. This method is very sensitive 
because stellar oscillation frequencies can be measured to extremely 
high precision, up to one part in 10$^9$ (Kepler et al.\ 2005). In the 
following it will be described what observables are available to base 
asteroseismic models on, how the measurements can be interpreted, and 
how observations and theory are used to develop methods for 
asteroseismic interpretations.

\section*{From the telescope to a seismic model}

\subsection*{Basic methods for analysing asteroseismic data}

Because stellar oscillations generate motions and temperature variations 
on the surface, they result in observable variability. The interplay of 
these variations causes light, radial velocity and line profile changes. 
Pulsating stars can thus be studied both photometrically and 
spectroscopically, via time series measurements.

These time series are subjected to {\it frequency analysis}, meaning that
the constituent signals are extracted from the data. In many cases, this
is done by harmonic analysis, transforming the time series into
frequency/amplitude space, e.g. by using the Discrete Fourier
Transformation

\begin{equation}
F_{N}(f) = \sum_{k=1}^{N}x(t_{k})e^{i2\pi ft_{k}},
\end{equation}

of the input function $x(t_{k})$, corresponding to the time series of 
the measurements. Periodograms, amplitude or power spectra are means of 
visualizing the results; an example is given in Fig.\ 5.

\begin{figure}[!htb]
\centering
\includegraphics[clip,angle=0,width=15.5cm]{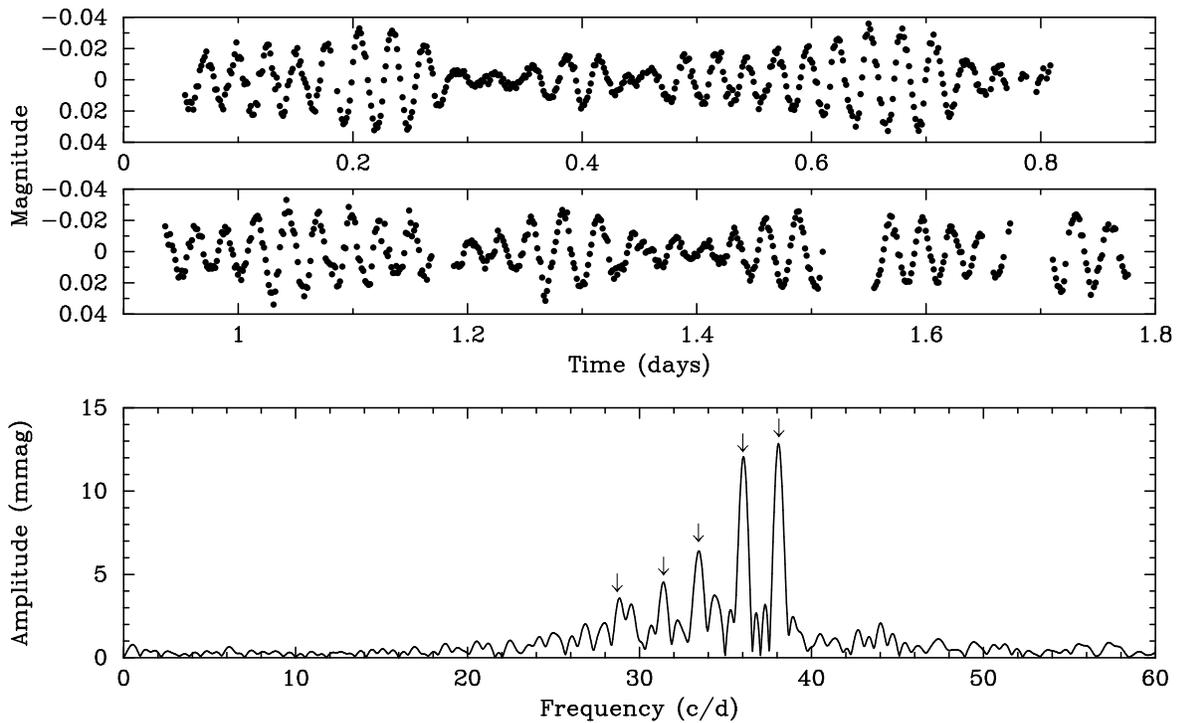}
\caption{The light variations (upper two panels) of a pulsating star and 
the corresponding Fourier amplitude spectrum. The complicated beating 
in the light curve is reflected by the presence of several signals in 
the periodogram; the strongest are labelled with arrows.}
\end{figure}

The amplitude spectrum in Fig.\ 5 can be used to estimate the 
frequencies of the dominant signals in the time series. In many cases, 
the analysis is carried forward by fitting sinusoids to the data, 
determining and optimizing their frequencies, amplitudes and phases, 
often by least squares methods. It has also become common practice to 
subtract this optimized fit from the data and to compute periodograms of 
the residuals, a procedure called {\it prewhitening}. This process is 
repeated until no more significant signals can be found in the data (a 
most delicate decision!). Depending on the specific case and 
requirements of the data set, a large number of alternative frequency 
analysis methods can also be applied, such as phase-dispersion 
minimization, autocorrelation methods, wavelet analysis etc. Care must 
be taken to keep the limitations of the data sets and of the applied 
methods in mind. Periods no longer than the data set itself can be 
reliably determined, and adjacent frequencies spaced no more than the 
inverse length of the data set can be resolved. Nonsinusoidal signals 
cause harmonics and combination frequencies in Fourier amplitude spectra 
that must not be mistaken for independent mode frequencies. No 
frequencies shorter than the inverse of twice the median distance 
between consecutive data points can be unambiguously detected; the 
highest frequency that can be retrieved in a given set of measurements 
is also called its {\it Nyquist frequency}.

Having determined the frequencies characterizing the stellar 
variability, the next step is their interpretation. Measurements of 
distant stars have an important limitation: nonradial oscillations 
create patterns of brighter and fainter, approaching and receding areas 
on the stellar surface. However, as a distant observer can usually not 
resolve the stellar surface, she or he can only measure the joint effect 
of the pulsations in light and radial velocity. As a consequence, the 
effects of oscillations with high spherical degree average out in 
disk-integrated measurements, and their observed amplitudes are reduced 
with respect to the intrinsic value. This effect is called {\it 
geometric cancellation} (Dziembowski 1977). Calculations show that the 
amplitude drops as $\approx 1/\sqrt{l}$ for high $l$. In ground based 
observational studies it is mostly assumed (and confirmed) that 
only modes with $l\leq2$ are observed in light and radial velocity, with 
a few exceptions of $l$ up to 4. Radial velocity measurements are 
somewhat more sensitive to higher $l$ than photometric observations.

Once the oscillation frequencies of a given star have been determined, 
how can they be made asteroseismic use of? As an example, high-precision 
radial velocity measurements of the close-by star $\alpha$ Centauri A 
showed the presence of solar-like oscillations; the power spectrum 
(square of amplitude vs.\ frequency) of these data is shown in Fig.\ 6.

\begin{figure}[!htb]
\centering
\includegraphics[clip,angle=0,width=15.5cm]{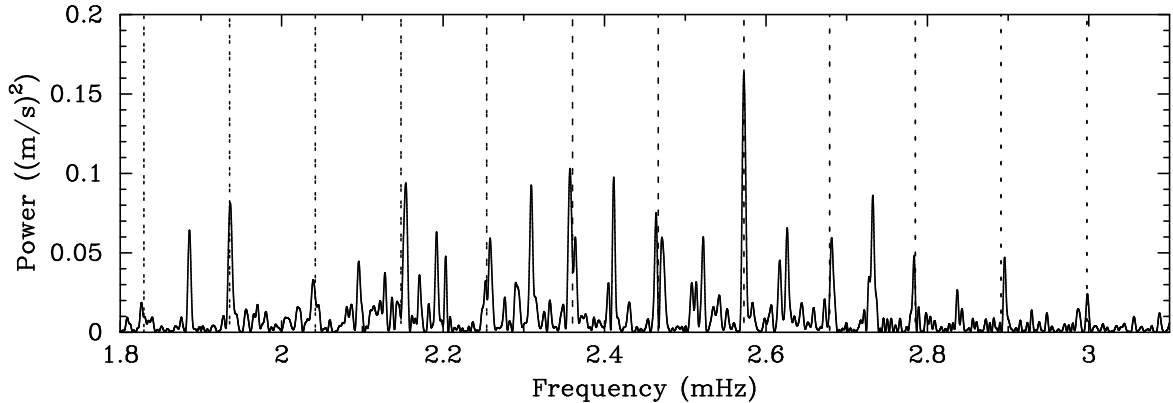}
\caption{High-overtone p mode spectrum of the solar-like oscillator 
$\alpha$ Centauri A (Bedding et al.\ 2004). The vertical dotted lines 
are separated by 106.2\,$\mu$Hz.}
\end{figure}

This graph contains a series of maxima equally spaced in frequency: a 
high-overtone p mode spectrum, as predicted by asymptotic theory. The 
mean frequency spacing is 106.2\,$\mu$Hz. However, it is obvious that 
only every other of the strongest peaks conforms to this spacing; there 
are other signals in between. The signals halfway between the vertical 
lines in Fig.\ 6, denoting modes of spherical degree $l$, are a 
consequence of Eq.~7: these are modes with $l\pm1$. Given the effects 
of geometrical cancellation, it is straightforward to assume that these 
are the modes of lowest $l$, viz. $l=0$ and 1.

However, there is more information present. A close look at Fig.\ 6 
reveals that many of the strongest peaks seem to be split. Again, the 
explanation for this finding lies within Eq.~7: the close neighbours 
are modes with $l+2$. These frequencies no longer have the same values 
as those with degree $l$ because the stellar interior has structure.

As stars evolve on the main sequence, their nuclear burning cores 
shrink, increase in density, and change in chemical composition as 
evolution progresses. This alters the acoustic sound speed in the core 
and is reflected in the frequency differences between modes of degrees 
$l, l+2$, also called the {\it small frequency separation 
$\delta\nu_{02}$}. On the other hand, 
evolution causes expansion of the outer regions of stars that become 
more tenuous, which increases the sound travel time through the star. 
This is measurable via the frequency difference between consecutive 
radial overtones and is called the {\it large frequency separation 
$\Delta\nu_0$}.

The large and the small separations can be computed for a range of 
theoretical stellar models. It turns out that a plot of $\delta\nu_0$ 
vs.\ $\Delta\nu_0$ allows an unambiguous determination of stellar mass 
and evolutionary state, and this method works particularly well for 
models with parameters similar to our Sun (main sequence, $M \simlt 1.3 
M_{\odot}$). This diagnostic is called an {\it asteroseismic HR Diagram} 
(Christensen-Dalsgaard 1988).

Another important tool to analyse high-overtone p mode pulsation spectra 
is the {\it Echelle Diagram}. This diagram plots the oscillation 
frequencies versus their modulus with respect to the large separation. 
Figure 7 shows the Echelle Diagram for $\alpha$~Cen~A constructed from 
the frequencies of the signals apparent in Fig.\ 6.

\begin{figure}[!htb]
\centering
\includegraphics[clip,angle=0,width=13.0cm]{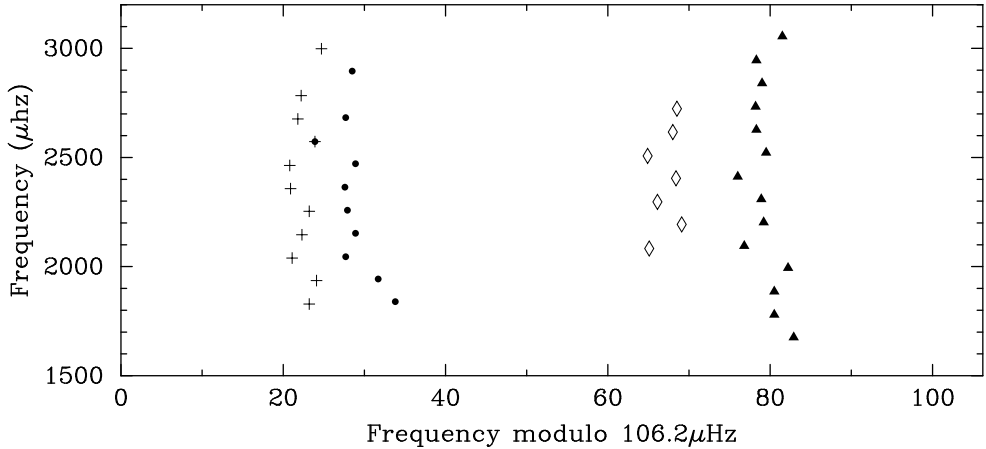}
\caption{Echelle diagram of the frequencies determined from Fig.\ 6 
(values from Bedding et al.\ 2004). Full dots represent radial modes, 
triangles mark dipole modes, plus signs stand for quadrupole modes and 
diamonds are for $l=3$ modes.}
\end{figure}

The frequencies fall onto four distinct ridges, corresponding to modes 
of the same spherical degree $l$. In this way, $l$ can be identified. 
From left to right, the ridges correspond to $l=2, 0, 3$ and 1, 
respectively, and it can be seen that, within the errors, 
$\delta\nu_{13}=5/3\delta\nu_{02}$, consistent with Eq.\ 7. The 
scatter in this diagram is due to a combination of the temporal 
resolution of the data and, more importantly, the finite lifetimes of 
the stochastically excited modes.

The example of $\alpha$~Cen~A shows that once a sufficient number of 
intrinsic pulsation frequencies of a given star has been determined, one 
may identify their spherical degrees by {\it pattern recognition}. This 
method is also applicable to the oscillation spectra of pulsating white 
dwarf stars, as exemplified in Fig.\ 8.

\begin{figure}[!htb]
\centering
\includegraphics[clip,angle=0,width=15.5cm]{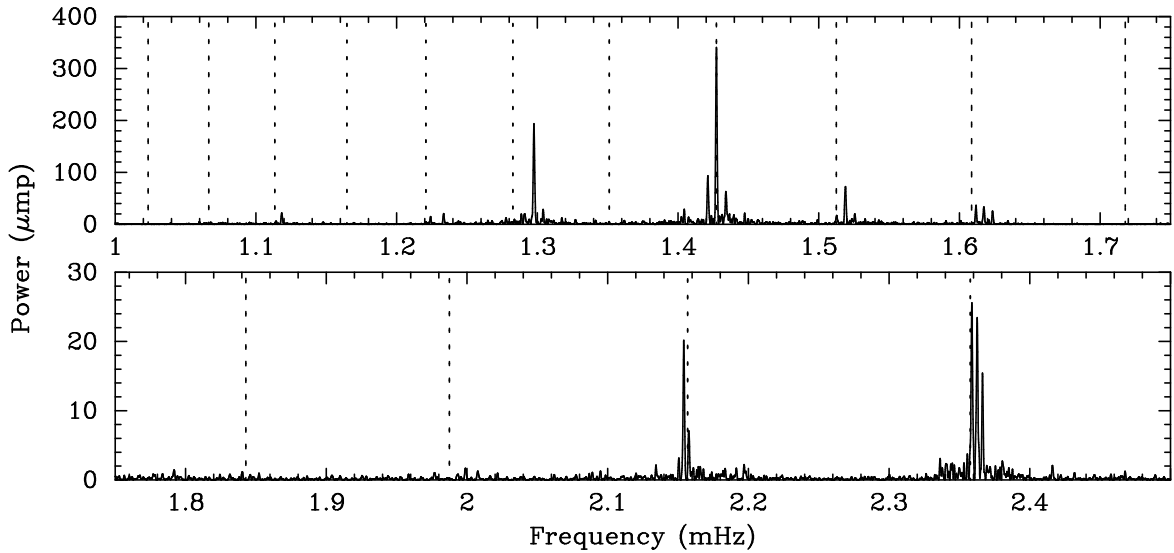}
\caption{High-overtone g mode spectrum of the pulsating DB white dwarf 
star GD~358 (Winget et al.\ 1994). The vertical dotted lines denote 
periods with an equal spacing of 39.5 seconds.}
\end{figure}

The series of peaks discernible in Fig.\ 8 does not form a pattern of 
equally spaced frequencies, but of equally spaced periods. This is the 
signature of a high overtone g mode pulsation spectrum (Eq.\ 8). In this 
case, all the strong peaks are roughly aligned with the vertical lines; 
there are no strong modes in between. Some of the expected signals are 
missing, but in most cases only apparently: they are just much weaker in 
amplitude. A mean period spacing of 39.5 seconds has been determined 
from this analysis, and the obvious identification of the strongest 
modes is $l=1$. Equation 8 then results in $P_0=55.9$\,s. For pulsating 
white dwarf stars, $P_0$ is a measure of the stellar mass, which was 
consequently determined.

Furthermore, several of the strongest peaks in Fig.\ 8 are split into 
triplets. These triplets are equally spaced in frequency, and they are 
the signature of rotationally split m modes (Eq.\ 9). The rotation 
period of GD 358 could therefore also be determined, just from an 
inspection of the power spectrum of its photometric time series, and 
basic application of theory. The fact that the $m$-mode splitting only 
results in triplets strengthens the previous identification of the 
spherical degree as $l=1$. The relative amplitudes within those triplets 
are dependent on the inclination of the stellar pulsation axis (Eq.\ 4), 
which may then be determined, but in reality there are other, presently 
unknown, effects that modify the relative multiplet amplitudes.

There is even more information present. It is noticeable that the peaks 
in the amplitude spectrum do not perfectly conform to their 
asymptotically predicted locations. This is a sign of {\it mode 
trapping}. White dwarf stars consist of a degenerate core, with subsequent 
outer layers of different chemical elements. The transition regions 
between these layers create spikes in the Brunt-Vais{\"a}l{\"a} 
frequency. The pulsation modes prefer to place their radial nodes in 
these transition regions, and be standing waves to both sides of the 
nodes. This modifies their frequencies compared to the case of 
homogeneous interior structure and gives rise to the observed deviations 
from equal period spacing.

The observant reader will have noticed three apparent inconsistencies in 
the previous paragraphs. How can two stars with similar oscillation 
frequencies be high-overtone p and g mode pulsators, respectively? How 
does one know that 106.2\,$\mu$Hz is the mean frequency spacing for 
$\alpha$~Cen~A, and not half this value? Why can the claimed small 
frequency separation not be the effect of rotation?

This is because some additional constraints are available that help in the
interpretation of pulsational mode spectra. As the p mode pulsation
periods depend on the sound travel time through the star, they must be
related to its size, or more precisely, to its mean density. The {\it
pulsation constant}

\begin{equation}
Q=P\sqrt\frac{\bar\rho_{\odot}}{\bar\rho},
\end{equation}

where $P$ is the pulsation period and $\bar\rho$ is the mean stellar
density, is a useful indicator of what type of mode one sees in a given
star. Over the whole HR diagram, the Q value for the radial fundamental
mode is between $0.03 - 0.04$\,d; for the Sun, it is 0.033\,d (0.8\,hr).
As mentioned before, radial modes can only be p modes, (pure) p modes
always have frequencies of the same order or larger than that of the
radial fundamental, and (pure) g modes always have frequencies lower than
that.

Because GD 358 is a white dwarf star and therefore has high mean 
density, its radial fundamental mode period would be of the order of 
4\,s. On the other hand, $\alpha$~Cen~A, a little more massive and more 
evolved than our Sun, has a radial fundamental mode period of about 
1\,hr. Therefore, the similar 5 - 10 minute pulsation periods of the two 
stars correspond to completely different types of mode.

The asymptotic p mode frequency separation also relates to the stellar
mean density

\begin{equation}
\Delta\nu_0=\Delta\nu_{0,\odot}\sqrt\frac{\rho_{\odot}}{\rho},
\end{equation}

where $\Delta\nu_{0,\odot}=135\mu$Hz. Given the knowledge on the mass 
and evolutionary state of $\alpha$~Cen~A, it immediately be inferred 
that 106.2\,$\mu$Hz must be the large frequency spacing. Finally, the 
rotation period of $\alpha$~Cen~A is much too long to generate $m$-mode 
splitting with the observed $\delta\nu_{02}$.

Apart from these examples, there is another clue towards the nature of 
observed pulsation modes just from observed frequency spectra. Radial 
modes are global oscillations that may reveal themselves because the 
period ratios of consecutive overtones are well known. On the main 
sequence, the period ratio of the radial fundamental mode and the first 
overtone is around 0.773, and the period ratio of the first to second 
overtone radial modes is 0.810. For more evolved stars, such as the 
$\delta$~Cephei stars, these ratios change to 0.705 and 0.805, 
respectively. Such period ratios are therefore suggestive to correspond 
to radial modes.

\subsection*{Methods for mode identification}

It is now clear that besides the pulsational mode spectra of stars 
themselves, the incorporation of other constraints is useful for their 
interpretation. This is particularly important for oscillation spectra 
that do not show obvious imprints of the underlying modes, like those of 
stars pulsating in high radial overtones. Examples are pulsations of low 
radial overtone or stars rotating so rapidly that the rotational 
splitting is of the same order as the frequency spacing of consecutive 
radial overtones of the same $l$ or of modes of the same $k$, but 
different $l$. In practice, most observed stellar oscillation spectra 
are incomplete, either because the star chooses so or because the 
observations do not have sufficient sensitivity, which makes it even 
more difficult to recognize patterns in the observed frequencies or 
periods and to type the modes accordingly. Therefore, {\it mode 
identification methods} have been developed, of which a variety is 
available.

The first method uses photometric data only. The flux change for 
nonradial pulsation in the linear regime can be expressed as

\begin{equation}
\Delta m (\lambda,t)=-1.086\epsilon P_l^{|m|}(\mu_0)((T_1+T_2)\cos(\omega 
t + \psi_T)+(T_3+T_4+T_5)cos(\omega t))
\end{equation}

(Watson 1988), where $\Delta m (\lambda,t)$ is the time and wavelength
dependent magnitude variation of an oscillation, $-1.086\epsilon$ is an
amplitude parameter transformed from fluxes to magnitudes, $P_l^{|m|}$
is the associated Legendre polynomial, $\mu_0$ is the cosine of the
inclination of the stellar pulsation axis with respect to the observer,
$\omega$ is the angular pulsation frequency, $t$ is time and $\psi_T$ is
the phase lag between the changes in temperature and local geometry
(mostly originating in convection zones near the stellar surface). The
term $T_1$ is the local temperature change on the surface, $T_2$ is the
temperature-dependent limb darkening variation, $T_3$ is the local
geometry change on the stellar surface, $T_4$ is the local surface
pressure change and $T_5$ is the gravity-dependent limb darkening
variation.

The $T_i$ terms can be determined for different types of pulsator from 
theoretical model atmospheres, and the observables best suited to reveal 
the types of mode present can also be deduced. These would for instance 
be the photometric amplitude ratios or phase shifts between different 
filter passbands, but also the optimal passbands themselves. An example 
of such a mode identification is shown in Fig.\ 9.

\begin{figure}[!htb]
\centering
\includegraphics[clip,angle=0,width=16.0cm]{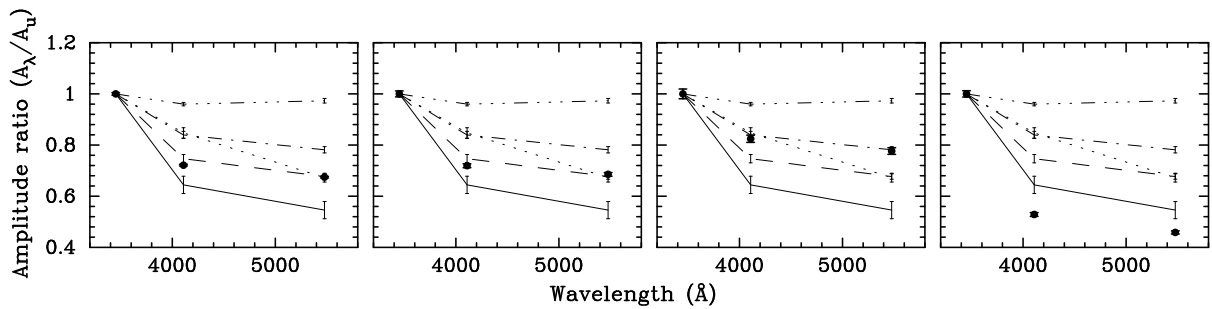}
\caption{Identification of the four strongest pulsation modes of the 
$\beta$~Cephei star 12~Lacertae from multicolour photometry (taken from 
Handler et al.\ 2006). The amplitudes are normalized to unity in the 
ultraviolet and compared with theoretical predictions. The full lines 
are for $l=0$, the dashed lines for $l=1$, the dashed-dotted lines for 
$l=2$, the dotted lines for $l=3$, and the dashed-triple dotted lines 
for $l=4$. The modes investigated in the two left panels are $l=1$, the 
next is $l=2$ and the rightmost one is $l=0$.}
\end{figure}

From photometry alone, only the spherical degree of a given pulsation 
mode can be identified. The method can be supported by adding radial 
velocity measurements, which increases its sensitivity 
(Daszy{\`n}ska-Daszkiewicz, Dziembowski, \& Pamyatnykh 2005), but still 
does not supply a determination of the azimuthal order. To this end, 
high resolution spectroscopy must be invoked.

The lines in a stellar spectrum are broadened by rotation through the 
Doppler effect: the intrinsic line profile is blueshifted on the parts 
of the stellar surface approaching the observer, and redshifted on the 
areas moving away. The effect is strongest on the stellar limb and 
decreases towards the centre. As a consequence, a rotationally broadened 
line profile contains spatial information of the stellar surface. For 
instance, in a dark starspot flux is missing, and a spike will move 
through the line profile as the spot rotates over the visible disk. The 
method to reconstruct stellar surface structure from line profile 
variations is called {\it Doppler Imaging} (Vogt \& Penrod 1983).

Apart from radial velocity changes, stellar oscillations also cause 
variations in rotationally broadened line profiles. The areas on the 
stellar surface that have an additional approaching component to their 
motion with respect to the observer have their contribution to the line 
profile blueshifted, whereas the receding parts are redshifted by the 
corresponding amount of pulsational Doppler velocity. The net result of 
all these motions is bumps travelling through the line profile, and 
their shapes are governed by the oscillation mode causing them (e.g., 
see Telting 2003 for a review). Examining stellar line profiles, 
pulsation modes of up to $l\approx20$ can be observed and identified, a 
vast extension in $l$ compared to photometric and radial velocity 
techniques. Some examples of pulsational line profile variations are 
shown in Fig.\ 10.

\begin{figure}[!htb]
\centering
\includegraphics[clip,angle=0,width=15.5cm]{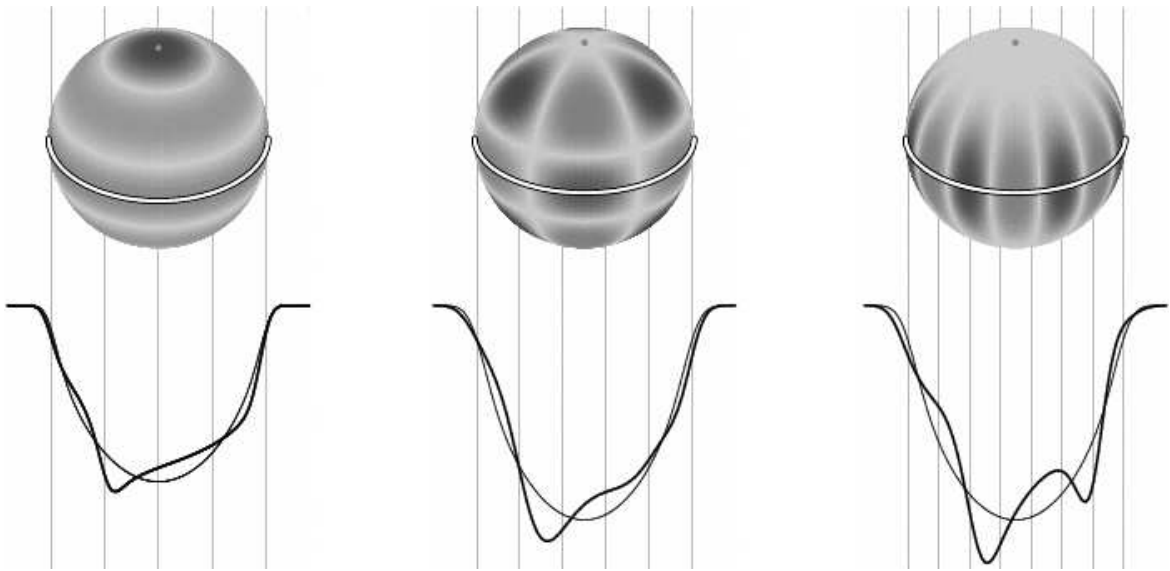}
\caption{Line profile variations due to stellar pulsation. The upper parts 
of the graph show the shape of the oscillation mode on the surface, 
whereas the thin lines in the lower halves represent the unperturbed 
rotationally broadened line profile, and the thick lines are the 
superpositions with the pulsation. Each individual mode (from left to 
right: $l=4, m=0; l=5, |m|=3; l=|m|=7$) generates a different distortion 
of the line profile. Adapted from Telting \& Schrijvers (1997).}
\end{figure}

The task now is to extract the correct values of $l$ and $m$ from the 
observed variations. The principle is to fit the theoretically 
calculated 3-D velocity field to the observed line profiles, and a wide 
range of spectroscopic mode identification methods is available. Some of 
the most commonly used are the Moment method (most suitable for low 
$l$), the pixel-by-pixel method, the Fourier Parameter Fit method, or 
Doppler reconstruction of the surface distortions. High-resolution 
spectroscopy is better suited for the determination of $m$ rather than 
$l$ which makes it complementary to photometric and radial velocity 
methods.

When resorting to photometric or spectroscopic methods, it is not 
required to arrive at unique identifications of all observed pulsation 
modes in each given star. What is needed is the secure identification of 
a sufficient number of modes to rule out all possible alternative 
interpretations. An example will be presented later.

\subsection*{Asteroseismic modelling}

The observed pulsation frequencies and their identifications are then 
matched to theoretical models (see Kawaler, this volume, for details).  
These would ideally be full evolutionary models with pulsation codes 
operating on them, although a few codes based on envelope models are 
still in use.

Most pulsation codes use the {\it linear approximation}: 
the oscillations are treated as linear perturbations around the 
equilibrium state. This allows the evaluation of the excitation of 
oscillation modes and thus the computation of theoretical domains of 
pulsation in the HR diagram. Nonlinear computations, that would allow 
predictions of oscillation amplitudes, are still rather the exception 
than the rule because they are, even today, expensive in terms of 
computing time, as are numerical hydrodynamical simulations.

Many stellar pulsation codes employ the {\it adiabatic approximation} 
(sometimes called isentropic approximation) to compute oscillation 
frequencies, which is the assumption that no energy exchange between the 
oscillating element and its surroundings takes place. Other codes 
perform nonadiabatic frequency computations. A wide variety of stellar 
oscillation codes is available, and most theory groups use their own 
routines, optimized for the application to their objects of main 
interest.

There are several strategies to find seismic models from observations. 
Some compare the observed oscillation frequencies with those of a grid 
of stellar models and perform automatic matching between them. This is 
computationally extensive, which means that supercomputers or parallel 
processing are invoked, or that intelligent optimization methods such 
as genetic algorithms are employed, or both.

Other strategies start with first imposing observational constraints. 
Besides the observed oscillation frequencies themselves and the 
identification of the underlying pulsation modes, these would often be 
estimates of the objects' positions in the HR diagram. As an
example, these are shown for the $\beta$~Cephei star $\nu$~Eridani in 
Fig.\ 11.

\begin{figure}[!htb]
\centering
\includegraphics[clip,angle=0,width=15.5cm]{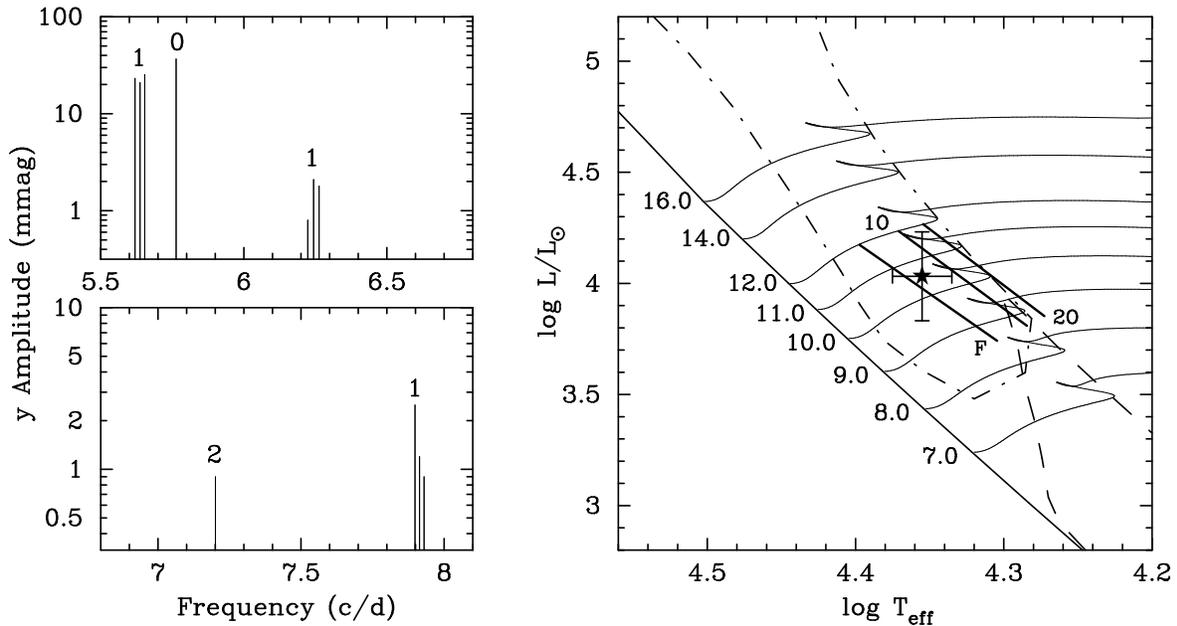}
\caption{Left: schematic oscillation spectrum of $\nu$~Eridani. The 
numbers on top of each mode (group) are their $l$ identifications, 
consistent in photometry and spectroscopy (De Ridder et al.\ 2004). 
Right: a plot of the star's position in the theoretical HR Diagram (star 
symbol) with its error bars and lines of equal mean density for the 
observed radial mode periods assuming they the fundamental, first and 
second overtones, respectively (thick lines). Some model evolutionary 
tracks labelled with their masses, the Zero-Age Main Sequence, and the 
borders of the $\beta$~Cephei (dashed-dotted line) and SPB star (dashed 
lines) instability strips are also shown.}
\end{figure}

The detection of a radial mode in the frequency spectrum is an asset for 
asteroseismic studies of this star: there are only three possibilities 
for the value of its mean density. These depend on whether its frequency 
corresponds to the fundamental, the first or the second radial overtone 
(Eq.\ 11). A comparison of the observed position of the star in the HR 
Diagram (right panel of Fig.\ 11) with its error bars leads to the 
rejection of the second overtone hypothesis, and to the elimination of 
models with masses below 8.5\,M$_{\odot}$.

Now the $l=1$ modes come into play: moving along the lines of constant 
mean density in the HR diagram, a comparison between their observed and 
theoretically predicted frequencies can be made. This is done on the 
left side in Fig.\ 12. Given the uncertainties and assumptions in the 
asteroseismic model construction, all $l=1$ modes are reproduced by 
models between $9.1 - 9.8$\,M$_{\odot}$ that have the radial mode as the 
fundamental. However, the frequency of the highest-overtone dipole mode 
cannot be explained within the errors for models that assume the 
first overtone for the radial mode. This interpretation can therefore 
also be rejected. The modal assignment to the observed frequencies is 
now unambiguous, and the range of models to be explored for seismic 
fitting is severely reduced.

The reason why this way of mode identification was successful is that 
the $g_1$ mode and its respective p mode neighbour are in the process of 
avoided crossing in the parameter space of interest. Small changes in 
the evolutionary states of the models depending on mass therefore lead 
to significant changes in their frequencies. The observed $l=2$ mode has 
been excluded from the mode identification procedure as all models in 
the parameter domain under consideration reproduce its frequency 
correctly: it is nearly a pure p mode and its frequency hardly changes 
in models of the same mean density.

\begin{figure}[!htb]
\centering
\includegraphics[clip,angle=0,width=15.5cm]{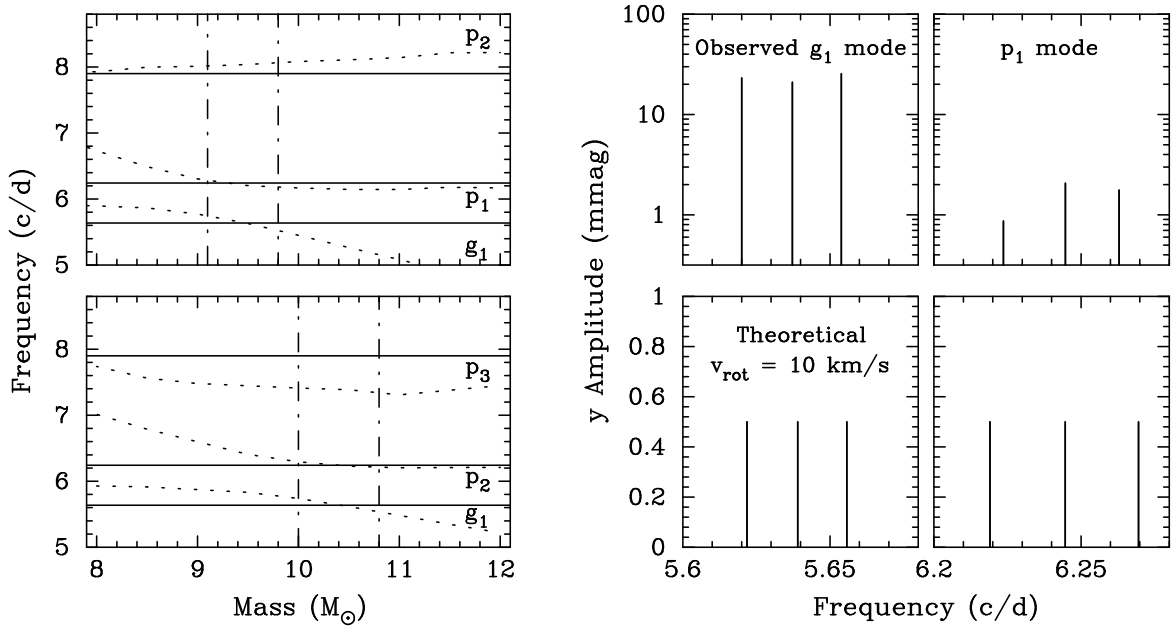}
\caption{Left: a figure showing the match of the $l=1, m=0$ modes of 
$\nu$~Eri for models of different mass but same mean density. The full
horizontal lines are the observed frequencies, the dotted lines are 
theoretical model frequencies. Upper panel: assuming that the radial 
mode is the fundamental. The vertical dashed-dotted lines show the mass 
range in which the first g mode and the first p mode fit the observed 
frequencies ($9.1 - 9.8$\,M$_{\odot}$). Lower panel: the same, but 
assuming the radial mode is the first overtone. Here it would be the 
first g mode and the second p mode that give an acceptable fit between 
$10.0 - 10.8$\,M$_{\odot}$. However, the observed $l=1$ mode with 
highest frequency is not compatible with the more massive models. Right: 
a comparison of the rotational splittings of a rigidly rotating model, 
with a rotation rate chosen to fit the observed $l=1, g_1$ triplet,
with the observations. The observed splitting of the $l=1, p_1$ triplet 
is not reproduced with this assumption, demonstrating the presence of 
differential interior rotation.}
\end{figure}

In this restricted parameter space, an interesting observation can be 
made: the observed $l=1$ m-mode splittings do not agree with those 
predicted by uniformly rotating models (Fig.\ 12, right-hand side). 
Fitting the rotational splitting for the $g_1$ mode, that samples the 
deep stellar interior, results in predicted splitting about 30\% larger 
than that observed for the $p_1$ mode, which is concentrated closer to 
the surface. This means that the star's rotation rate increases towards 
its interior, as predicted by theory (Talon et al.\ 1997). In addition, 
as the $l=1, g_1$ mode frequency is sensitive to the size of the 
convective core (it is the $g_c$ mode!), a constraint on the convective 
core overshooting parameter was also obtained (Pamyatnykh, Handler, \& 
Dziembowski 2004)

This example shows the potential of asteroseismology, even if only few 
pulsation modes are available. Radial pulsations are most valuable 
in the identification process as they immediately provide accurate 
constraints on the stellar mean density that can become unambiguous 
if the radial overtone of the mode can be inferred, supported by other 
estimates such as on effective temperature and luminosity. Theory and 
observations work hand in hand. Once a unique identification of the 
normal modes is achieved, firm results on stellar structure can be 
obtained, even given uncertainties in the modelling procedure.

In the case of $\nu$~Eridani, the most severe problem is that the $l=1, 
p_2$ mode is hard to be excited and accurately matched in frequency 
given the observed effective temperature of the star and its chemical 
surface composition. Only changes to the input physics would solve this 
problem. This is just the goal of asteroseismology: to improve our 
knowledge about stellar physics. Such an improvement can only be 
achieved using observational results that present models cannot account 
for!

\section*{Applications}

\subsection*{Pulsating white and pre-white dwarf stars}

The first pulsators that could be studied asteroseismically were white 
dwarf and pre-white dwarf stars, pulsating in high-overtone low-degree g 
modes. A remarkable initial result was the theoretical prediction of a 
then new class of pulsating white dwarfs of spectral type DB, and their 
subsequent observational confirmation (Winget et al.\ 1982). Upon the 
realization that some pulsating white dwarf stars have very complicated 
frequency spectra, and having a well-developed theory of white dwarf 
pulsation available, the main obstacle to extracting correct mode 
frequencies from observational data was the daily interruptions of the 
measurements by sunlight. Figure 13 explains why.

\begin{figure}[!htb]
\centering
\includegraphics[clip,angle=0,width=14.5cm]{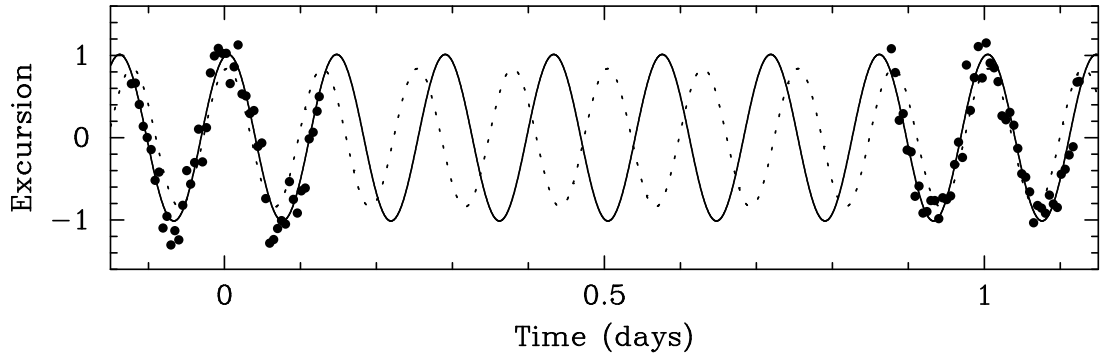}
\caption{Simulation of observations from a single astronomical site. The 
filled circles represent the measurements, including noise. The full 
line is a fit with the correct frequency present in the data, and the 
dotted line is a fit with a frequency different by 1 cycle per day. The 
two fits represent the data almost equally well. However, they would be 
completely out of phase, and therefore easily separable, if measurements 
were available in between the present data, from a site 180 degrees 
different in geographical longitude. This led to the setup of multisite 
telescope networks.}
\end{figure}

The interruptions cause ambiguities in the determination of the 
frequencies of the signals present in the data: a frequency different by 
one cycle per sidereal day from the real oscillation frequency would 
generate a fit of comparable quality. In the presence of complicated 
variability and most notably for signals of low signal-to-noise this 
could lead to erroneous frequency determinations. Any seismic model 
based on incorrect observational input is misleading.

The solution to this problem is to avoid, or at least minimize, daytime 
gaps in time resolved measurements. This can be accomplished by 
concerted observational efforts, involving interested colleagues over 
the whole globe, passing on the asteroseismic target from one 
observatory to the next. The best known of these collaborations is the 
Whole Earth Telescope (WET, Nather et al.\ 1990), invented for the study 
of pulsating white dwarf stars.

One of the first WET runs was devoted to the prototypical pulsating 
pre-white dwarf star PG~1159-035 = GW~Vir. It resulted in the detection 
of over 100 g-mode frequencies of $l=1$ and 2 modes of high radial 
overtone, leading to precise determinations of the stellar mass and 
rotation period, an asteroseismic detection of compositional 
stratification, and an upper limit to the magnetic field strength 
(Winget et al.\ 1991). Subsequent WET observations of the prototypical 
pulsating DB white dwarf star GD~358 = V777~Her showed a mode spectrum 
dominated by high-overtone $l=1$ g~modes (cf.\ Fig.\ 8), resulting in 
determinations of its total and surface Helium layer mass, luminosity 
and rotation rate (Winget et al.\ 1994).

The evolution of white dwarf stars is dominated by cooling, at (nearly) 
constant radius. As they cool, they pass through a number of instability 
strips. Evidence is that all white dwarf stars located in such 
instability domains in the HR diagram do pulsate. This has an important 
consequence: the interior structures of the pulsators must be 
representative of all white dwarf stars. Thus asteroseismic results for 
white dwarf pulsators can be extended to the stars that do not 
oscillate.

Cooling of pulsating white dwarf stars changes their oscillation 
periods, and the rate of period change is directly related to their 
energy loss. The hottest DB pulsators have reasonably high neutrino 
emission rates, and their evolutionary period changes may be able to 
tell us whether their neutrino emission is compatible with the standard 
model of particle physics (Winget et al.\ 2004). Measurements to detect 
such a period change are ongoing. On the other hand, the period changes 
of DA pulsators (some of which are the most stable optical clocks in the 
Universe, Kepler et al.\ 2005) could reveal the masses of axions, if 
the latter existed (Kim, Montgomery, \& Winget 2006).

Asteroseismology of white dwarf pulsators does not only allow to detect 
stratification in their chemical profiles near the surface, it also 
gives evidence of their core composition. This, in turn, is dependent on 
their history of evolution on the Asymptotic Giant Branch (AGB), and can 
be used to obtain constraints on the nuclear reaction rates in AGB 
stars. Present results (Metcalfe 2005) indicate consistency with 
evolutionary models.

As white dwarf stars cool, their cores become crystallized. Being 
composed mainly of carbon and oxygen, such cores can be seen as giant 
diamonds! Massive white dwarf stars begin to crystallize when still in 
the DAV (ZZ Ceti) instability strip, and asteroseismic investigations of 
one such massive pulsator have proven substantial crystallization in its 
interior (Metcalfe, Montgomery \& Kanaan 2004, Brassard \& Fontaine 2005).

The light curve shapes of pulsating white dwarf stars of spectral types 
DB and DA are often nonsinusoidal. The nonlinearities originate in their 
convection zones, that cannot instantly adjust to the pulsational 
variations (the g mode pulsations of pulsating white dwarf stars are 
almost exclusively due to temperature changes; Robinson, Kepler, \& 
Nather 1982). As the light curve shapes of such pulsators depend on the 
thermal response time of the convection zone, the latter parameter can 
be determined from nonlinear light curve fits (Montgomery 2005).

As a final example, there are pulsating white dwarf stars in mass 
accreting close binary systems. If the mass transfer rate is in a 
certain range, the surface temperature of the accreting white dwarf 
places it in an instability strip. A handful of such oscillators is 
known to date (Mukadam et al.\ 2007), but attempts at asteroseismology 
have proven difficult due to low amplitudes and unstable mode spectra.

Among all classes of pulsating star (aside from the Sun itself), 
asteroseismology of pulsating white dwarf stars is certainly in the most 
advanced state. A new class of such oscillators has recently been 
proposed, hot DQ stars with carbon-dominated atmospheres and 
temperatures similar to that of the DB pulsators (Montgomery et al.\ 
2007, Dufour et al.\ 2009). There is little doubt remaining that the 
variability of these stars is due to pulsation.  Theory predicts yet 
another new type of white dwarf oscillator, DA white dwarf stars 
somewhat hotter than the DB pulsators, but observational searches for 
them have so far been inconclusive (Kurtz et al. 2008). We refer to 
Montgomery (2009), and references therein, for more information on 
asteroseismology of pulsating white dwarf stars.

\subsection*{Delta Scuti stars}

At about the same time when the necessity for worldwide observing 
efforts for pulsating white dwarf stars was realized, the same 
conclusion was reached for $\delta$~Scuti stars, some of which also 
exhibit complex oscillation spectra composed of g, p and mixed modes of 
low overtone. The Delta Scuti Network, founded over 25 years ago, was 
the first multisite observing collaboration for these stars, followed by 
a number of others such as STEPHI or STACC; a few WET runs were also 
devoted to $\delta$~Scuti stars.

The asteroseismic potential of $\delta$~Scuti stars is enormous, but 
could so far not be fully exploited. Part of the reason is visible in 
Fig.\ 4: the pure p mode spectrum on the ZAMS mainly allows a 
determination of the stellar mean density. When the scientifically more 
interesting mixed modes appear, the frequency spectrum is fairly dense 
and requires a large amount of data and long time base to be resolved 
observationally. However, the real stars seldom co-operate in showing 
many of the potentially excited modes at observable amplitude, 
inhibiting mode identification by pattern recognition. In addition, mode 
amplitudes are often small, counteracting reliable identifications of 
many modes by applying the methods discussed before. Even though dozens 
or even hundreds of pulsation modes have been detected in some 
$\delta$~Scuti stars, little has been learnt on their interior 
structures from asteroseismology so far.

Pulsational amplitude limitation of $\delta$~Scuti stars is a major 
problem for theory: what makes the star excite only certain modes, which 
modes would these be and what determines their amplitudes? Observational 
evidence suggests that in evolved stars mode trapping is (part of) the 
answer, and that oscillations with frequencies around those of radial 
modes are preferentially excited (Breger, Lenz, \& Pamyatnykh, 2009).

For slowly rotating $\delta$~Scuti stars gravitational settling and 
radiative levitation give rise to chemical surface peculiarities and are 
believed to deplete the pulsational driving regions. Consequently, Am 
and Ap stars are not expected to pulsate, although a few of them do 
(Kurtz 2000). This also means that many $\delta$~Scuti stars rotate 
rapidly, which requires special calculations to extract information from 
their distorted pulsation modes, a field that has made 
considerable progress in the recent past (Reese et al.\ 2009).

Some pre-main sequence stars cross the $\delta$~Scuti instability strip 
on their way to the ZAMS, and consequently pulsate. The interior 
structures of these stars are fairly simple, which may make them more 
accessible to asteroseismic investigation compared to their main 
sequence counterparts. The oscillation spectra of pre-main sequence and 
main sequence $\delta$~Scuti stars in the same position in the HR 
Diagram are predicted to be different, which may allow an asteroseismic 
separation (Suran et al.\ 2001).

\subsection*{Slowly Pulsating B and Gamma Doradus stars}

These two classes of high-overtone g mode pulsator, although well 
separated in effective temperature, share most of their asteroseismic 
characteristics. They also share the problems with respect to 
observations, caused by their long periods: resolving their oscillation 
spectra requires measurements over a long time baseline, possibly many 
years. It is therefore no surprise that most of the known members of 
these two groups of pulsator were discovered with the HIPPARCOS (HIgh 
Precision PARallax COllecting Satellite) mission from its data set 
spanning over three years (Waelkens et al.\ 1998, Handler 1999).

As in pulsating white dwarf stars, effects of inhomogeneities in stellar 
structure would manifest themselves in mode trapping and thus in
oscillatory behaviour in the g mode periods  
(Miglio et al.\ 2008). The dominant inhomogeneity is the change 
in mean molecular weight at the edge of the convective core, whose size 
can be measured. This is a method alternative to measuring the frequency 
of the $g_c$ mode in p mode pulsators.

Several of the SPB and $\gamma$ Doradus stars rotate with periods 
comparable to their oscillation periods. This, again, calls for models 
that take rotation into account with a more sophisticated approach than 
perturbation theory; the corresponding work is in progress.

The $\gamma$ Doradus stars are located in a domain where the influence 
of convection on the pulsations is significant; convection is also 
responsible for the red edge of the $\delta$~Scuti instability domain. 
Modelling with a time dependent convection approach allowed Dupret et 
al.\ (2005) to reproduce the observed boundaries of these instability 
strips, and also to perform predictions of mode excitation in these 
stars. The situation for the SPB and $\gamma$~Doradus stars with respect 
to asteroseismology is therefore the same as for $\delta$~Scuti stars: 
the basic theory is in place, the difficulty remains to find stars 
permitting the extraction of the required information from the 
observations.

\subsection*{Beta Cephei stars}

The $\beta$~Cephei stars are massive ($\sim9-17 M_{\odot}$) early-B main 
sequence stars that oscillate radially and nonradially in p, g and mixed 
modes of low radial overtone. This is roughly the same type of modes as 
excited in the $\delta$~Scuti stars, but asteroseismology has been more 
successful for $\beta$~Cephei stars in the recent past due to several 
reasons.

The observed frequency spectra are simple enough to provide initial 
clues for mode identification, yet complicated enough to reveal 
information about the stars' interior structures. Photometric and 
spectroscopic mode identification methods (and combinations of both) 
could be applied successfully to some $\beta$~Cephei stars (e.g., De 
Ridder et al.\ 2004), additionally aided by the large radial velocity to 
light amplitude ratios (of the order of several hundreds km/s/mag) of 
their pulsation modes. Sufficient information for unique identifications 
of all modes was obtained; an example was shown earlier.

Asteroseismic modelling was also eased for $\beta$~Cephei stars, as 
radial modes have sometimes been identified; an example was shown 
earlier. This immediately reduces the parameter space in which a seismic 
model must be sought by one dimension. Due to the evolutionary state of 
$\beta$~Cephei stars (near the centre of the main sequence), a few of 
the observed nonradial modes are of mixed p/g type, which has provided 
information about the convective core size and/or differential interior 
rotation in a number of stars (e.g., Aerts et al. 2003, Pamyatnykh et 
al. 2004).

Apart from learning about stellar interiors, asteroseismology of 
$\beta$~Cephei stars has interesting astrophysical implications. Given 
their high masses, they are progenitors of supernovae of Type II which 
are largely responsible for the chemical enrichment of galaxies. The 
evolution of massive stars is strongly affected by rotational mixing and 
angular momentum transport (Maeder \& Meynet 2000); their internal 
rotation profile is testable by asteroseismology of $\beta$~Cephei stars.

However, the field has not yet matured to a point where we can claim 
satisfactory understanding of all aspects of the physics governing the 
interior structures of massive main sequence stars. Several questions 
may be answered by seismic sounding of $\beta$~Cephei stars: 
how strong is differential interior rotation? How efficient is 
internal angular momentum transport (Townsend 2009)? How strong is 
convective core overshooting? Can all stars between $\sim9-17 M_{\odot}$ 
be modelled with the same convective overshooting parameter?

There are additional questions related to the pulsation physics of 
$\beta$~Cephei stars that need to be addressed. Only the centre of the 
theoretically predicted $\beta$~Cephei instability strip is populated by 
observed pulsators. Is this an observational shortcoming or a weakness 
of theory? What is the upper mass limit of the $\beta$~Cephei stars? Are 
there post-main sequence $\beta$~Cep stars, contrary to theoretical 
predictions?

Five different observables strongly depend on the opacities and element 
mixtures used for theoretical modelling of $\beta$~Cephei stars: the 
radial fundamental to first overtone period ratio, the excited range of 
pulsation modes, the frequencies of p modes with radial overtones larger 
than two, the dependence of bolometric flux amplitude on the surface 
displacement (Daszy{\'n}ska-Daszkiewicz \& Walczak 2009) and in case of 
a ``hybrid" pulsator (see below) the excited range of g~modes. Most of 
these observables are largely independent of each other; modelling of 
some stars shows that no standard input opacities and element mixtures 
can explain the pulsation spectra in detail. Therefore, the last 
question that may be answered from asteroseismology of $\beta$~Cephei 
stars is: where must we improve the input physics for stellar modelling?

Some $\beta$~Cephei and SPB stars have emission-line spectra and are 
thus Be stars. These objects rotate rapidly and have circumstellar 
disks, occasional outbursts etc. They may also be studied 
asteroseismically, but their oscillations are hard to be 
detected and identified, and their rapid rotation requires special 
theoretical treatment - that is underway.

\subsection*{Pulsating subdwarf stars}

Three types of pulsating subdwarf star are known: long-period subdwarf 
B (sdB) stars that oscillate in high-overtone g modes (V1093 Herculis 
stars), short-period sdB stars pulsating in low-overtone p and g modes 
(V361 Hydrae stars), and the only oscillating subdwarf O (sdO) star 
known to date, a low-overtone p mode pulsator.

Although their g modes would allow the sounding of deep interior 
regions, their faintness, long periods and low amplitudes made V1093 
Herculis stars escape asteroseismic study so far. Theoretical studies of 
the sdO pulsator have so far been focused on the problem of mode 
excitation. Therefore, the only subdwarf pulsators that have been 
asteroseismically modelled are among the V361 Hydrae stars.

This is no easy undertaking as the problem with mode identification and 
mostly sparse (but for a few stars rich and highly variable) frequency 
spectra again occurs. It is possible that these objects have steep 
interior rotation gradients. Pulsation models must also be built upon 
evolutionary models including the effects of gravitational settling and 
radiative levitation. In practice, modelling is carried out by surveying 
parameter space in effective temperature, gravity, mass and hydrogen 
mass fraction and by seeking best agreement between observed and 
theoretically predicted oscillation frequencies. Results have been 
obtained on about a dozen of those stars, and a mass distribution 
consistent with that expected from a double star evolutionary scenario 
has been obtained. Charpinet et al.\ (2009) elaborated on many aspects 
of asteroseismic modelling of pulsating subdwarf stars.

\subsection*{Rapidly oscillating Ap stars}

The rapidly oscillating Ap (or short: roAp) stars are special among the
pulsating stars because their high-overtone p mode oscillations are
predominantly governed by a magnetic field that aligns their pulsation
axis with the magnetic, and not with the rotation, axis of the stars. The 
magnetic field also distorts the pulsations modes, so they can no longer 
be described with a single spherical harmonic.

Because of their short periods and very low amplitudes (below 1\%) these 
stars have for many years been studied photometrically only; an 
extensive review was given by Kurtz \& Martinez (2000). Spectroscopic 
observations of roAp stars, however, provided a new level of insight, 
particularly into atmospheric physics of these pulsators.

The reason is that the vertical wavelength of the pulsation modes is 
shorter than or about the same order of the layer thicknesses of the 
chemical elements in their atmospheres that are highly stratified by 
radiative levitation. Therefore the radial velocity amplitudes of the 
oscillations change with line depth and from one chemical species to the 
other. The chemical elements are also inhomogeneously distributed over 
the surface, allowing three-dimensional studies of the abundances and 
pulsational geometry (Kurtz 2009).

Because of the unique possibilities offered by the atmospheric structure 
of roAp stars, spectroscopy is also much more sensitive in detecting 
oscillations compared to photometry. As an outstanding example, 
Mkrtichian et al.\ (2008) detected a complete $l=0-2$ mode spectrum for 
Przybylski's star over $3-4$ radial orders and performed some initial 
seismic modelling.

\subsection*{Solar-like oscillators}

The low amplitudes of the stochastically excited oscillations of 
solar-type stars made their observational detection elusive for a long 
time. In retrospect, the first detection was made by Kjeldsen et al.\ 
(1995), but confirmed only several years later. Meanwhile, the 
observational accuracy has improved to an extent that detections were 
made in hundreds of stars (mostly giants), and seismic analyses of 
several were performed.

The potential of solar-like oscillations for seismic sounding is large. 
Once detected, the pulsation modes are rather easy to be identified 
because they are high-overtone p modes; an example was provided earlier. 
So are the large and the small frequency separations (if detectable), 
immediately placing main sequence stars on the (asteroseismic) HR 
diagram.

Most interesting, as for all pulsators with nearly-asymptotic frequency 
spectra, are irregularities in the latter, caused by features in the 
stellar interiors. These would for instance be the base of their 
envelope convection zones, or the helium ionization region. Houdek 
(2009) gave an overview of the expected seismic signatures of such 
features and their astrophysical importance.

The phenomenon of avoided crossing (Fig.\ 4) also takes place in 
solar-like oscillators once they have reached the subgiant stage, and it 
makes itself obvious in Echelle Diagrams (Bedding et al.\ 2007). As for 
other types of pulsator, this would allow an asteroseismic determination 
of the convective core size (for stars sufficiently massive to possess a 
convective core).

The limited lifetimes of the intrinsically damped solar-like 
oscillations enable inferences concerning pulsation mode physics. The 
observed power spectra at individual mode frequencies show a multitude 
of peaks whose overall shape would correspond to a Lorentzian. The 
half-widths of these Lorentzians yield a determination of the mode 
damping rates; the {\it mode lifetimes} are inversely proportional to those. 
The mode lifetimes are in 
turn dependent on properties of the surface convection zone.

Theoretical predictions of the amplitudes of solar-like oscillators are 
important not only for understanding their physics, but also for 
planning observational efforts. After years of predictions resulting 
in amplitudes larger than were observed, the incorporation of the mode 
lifetimes and subsequent computation of {\it mode heights} appear to 
result in a scaling law that estimates observed amplitudes well (Chaplin 
et al.\ 2009), and seems in agreement with measurements up to 
oscillating giants (Hekker et al.\ 2009).

Finally, it is worth to note that as all cool stars possess a convective 
envelope, it can be expected the solar-like oscillations are excited in 
all of them, up to red supergiants (Kiss et al.\ 2006).

\subsection*{Hybrid pulsators}

Some of the pulsational instability strips shown in Fig.\ 1 partly 
overlap. It is therefore logical to suspect that stars that belong to 
two different classes of pulsating star, having two different sets of 
pulsational mode spectra excited simultaneously, may exist. Indeed, a 
number of those have been discovered. This is good news for 
asteroseismology as the information carried by both types of 
oscillation can be exploited. The confirmed cases have so far always 
been high-overtone g modes and low-overtone p modes, as evaluations of 
the pulsation constants of the oscillations show. Mixed-mode pulsations 
by themselves are not "hybrid" pulsations because they occupy the same 
frequency domain as pure p~modes.

There are $\delta$~Scuti/$\gamma$~Doradus stars, $\beta$~Cephei/SPB 
stars and long/short-period subdwarf B pulsators. The main physical 
difference between the B-type and A/F-type "hybrid" pulsators is that in 
the first group the same driving mechanism excites both types of 
oscillation, whereas in the $\delta$~Scuti/$\gamma$~Doradus stars two 
main driving mechanisms are at work. The cooler $\delta$~Scuti stars and 
all $\gamma$~Doradus stars should have thin surface convection zones 
that would support the excitation of solar-like oscillations (e.g., 
Samadi et al. 2002). Ongoing searches for such oscillations have so far 
remained inconclusive, but there has been a report of solar-like 
oscillations in a $\beta$~Cephei star (Belkacem et al.\ 2009).

The revision in our knowledge of the solar chemical element mixture 
(Asplund et al. 2004) resulted in the prediction of a much larger 
overlap region between $\beta$~Cephei and SPB stars in the HR diagram, 
and the frequency ranges of the excited long and short period modes in 
"hybrid" B-type pulsators suggest that the heavy-element opacities used 
for stellar model calculations are still too low (Dziembowski \& 
Pamyatnykh 2008, Handler et al.\ 2009). Intriguingly, a similar 
conclusion has been independently obtained from helioseismology (Guzik, 
Keady, \& Kilcrease 2009).

One of the first hybrid pulsators reported in the literature (HD 209295, 
Handler et al.\ 2002) turned out to have its g mode oscillations tidally 
excited by a another star in a close eccentric orbit. The changing 
gravitational influence of the companion gives rise to forced 
oscillations with frequencies that are integer multiples of the orbital 
frequency. The tidal deformation of the pulsating star is similar to a 
sectoral $l=2$ mode, and such modes are therefore most easily excited. 
There are also several known cases of pulsation in ellipsoidal 
variables. Whereas close binarity represents an additional complication 
for theoretical asteroseismic studies due to the gravitational 
distortion of the pulsator, some other cases of binarity can be used to 
obtain additional constraints for seismic modelling.

\subsection*{Pulsation in eclipsing binaries and open clusters}

The fundamental way to determine stellar physical parameters, in 
particular masses, is the analysis of detached eclipsing binary systems 
whose components can be assumed to have evolved as if they were single 
(Torres et al.\ 2010). Such constraints are most welcome for 
asteroseismology and therefore it is self-evident that pulsators be 
sought for in such binaries. To date, several dozens of such systems are 
known. Most of these would be $\delta$~Scuti pulsators, some are 
$\beta$~Cephei stars, but the best studied case is a short-period sdB 
pulsator.

These objects provide another possibility for mode identification 
(Nather \& Robinson 1974): throughout the eclipse, different parts of 
the stellar surface become invisible. In case of nonradial oscillations, 
only part of the pulsation modes is seen, and the light amplitudes and 
phases change according to the types of mode. In this way, the 
oscillation mode can be identified, a method known as {\it eclipse 
mapping} or {\it spatial filtration} (e.g., Reed, Brondel, \& Kawaler 
2005).

Another way to gain support for asteroseismic modelling is to study 
pulsators in stellar clusters. Besides the possibility to observe 
several objects simultaneously (e.g. via CCD photometry or multifibre 
spectroscopy), cluster stars can be assumed to have originated from the 
same interstellar cloud. Therefore, they should be of the same age and 
chemical composition. These parameters can be well determined from the 
properties of the cluster itself, and be imposed as a constraint on the 
seismic modelling procedure of all pulsating cluster members, also known 
as {\it ensemble asteroseismology}.

\subsection*{A new era in precision}

Most observational results reported in this article so far were based on 
classical ground based observing methods, such as single- and multisite 
photometry and spectroscopy. However, a new era in observational 
asteroseismology has begun.

Asteroseismology requires knowledge about as many intrinsic stellar 
oscillation frequencies as possible, and these often have low amplitude, 
calling for measurements with the highest accuracy. As this is also a 
requirement for the search for extrasolar planets, synergies between the 
two fields have emerged.

Spectroscopically, asteroseismology has benefited from high precision 
radial velocity techniques, such as the iodine cell method, originally 
invented to find extrasolar planets. Measurements of oscillations in 
distant stars with amplitudes down to 20 cm/s, about one-tenth of human 
walking speed, have become possible in that way. Only about a dozen 
spectrographs worldwide are capable of reaching the required precision. 
Such observations are still therefore expensive in terms of observing 
time and complexity of data analysis, but new observing networks, such 
as the Stellar Observations Network Group (SONG, 
http://astro.phys.au.dk/SONG/) aim at achieving similar precision on a 
regular basis. SIAMOIS (Sismom{\`e}tre Interf{\'e}rentiel {\`A} Mesurer 
les Oscillations des Int{\'e}rieurs Stellaires, Mosser et al.\ 2009) is 
expected to work at the same precision and duty cycle with its single node 
placed in Antarctica.

Concerning photometry, the main problem of ground based observations is 
scintillation, irregular changes in the measured intensity due to 
anomalous atmospheric refraction caused by turbulent mixing of air with 
different temperatures. The solution to this problem is to observe 
stellar variability from space. Here the synergy with extrasolar planet 
research is that the measurements have sufficient precision to detect 
the signature of transits of planets when they pass in front of their 
host star.

Aside from fine-guidance sensor Hubble Space Telescope photometry, the 
first asteroseismic data from space were due to an accident. 
The main science mission of the WIRE (Wide-Field Infrared Explorer) 
satellite failed due to loss of cooling fluid for the main instrument, 
but the star trackers, of 52 mm aperture, were consequently used 
for time-resolved photometry. An overview of the results can be found in 
the paper by Bruntt \& Southworth (2008); WIRE has fully ceased 
operation in October 2006.

The first dedicated asteroseismic space mission that was successfully 
launched is MOST (Microvariability and Oscillations of STars, 
http://www.astro.ubc.ca/MOST/, Walker et al.\ 2003, Matthews 2006). The 
spacecraft is in orbit since June 2003 and still continues to provide 
asteroseismic data with its 15-cm telescope as main instrument. One of 
the most interesting results from MOST was the discovery of 
high-overtone g mode oscillations in a supergiant B star (Saio et al.\ 
2006), suggesting an extension of the SPB star instability domain to the 
highest luminosities.

The CoRoT (Convection, Rotation and Transits, Baglin 2003) mission was 
successfully launched in December 2006 and hosts a 27-cm telescope. It 
observes two fields of $2\times3$ degrees on the sky each, one devoted 
to asteroseismology of a few bright stars, and the other searching for 
planetary transits in many stars, at the same time performing a 
high-precision stellar variability survey. A special volume of Astronomy 
\& Astrophysics (2009) reports some of the early CoRoT science. 

Asteroseismically, the detection of solar-like oscillations in almost a 
thousand giant stars is remarkable (Hekker et al.\ 2009). In addition, 
one of the most intriguing CoRoT results was the detection of large 
numbers of variability frequencies in two $\delta$~Scuti stars. 
Interpreted as independent pulsation modes, these would supply hundreds 
of oscillations to be asteroseismically modelled, reverting the basic 
problem for the study of these objects: first, there were too few modes 
available, now there would be too many! However, doubts have been raised 
whether all the frequencies extracted from those data would really 
correspond to normal modes of pulsation, or would rather be a signature 
of granulation (Kallinger \& Matthews 2010).

Originally designed for the search for earth-like planets in the habitable
zone, the latest addition to the asteroseismic space fleet became Kepler
(http://www.kepler.nasa.gov), launched in March 2009. Asteroseismology can
measure the radii of solar-like oscillators, among which should be
planet-hosting stars, to a relative accuracy of 3\%. Thus the radii of
transiting planets would be known to the same precision. Therefore it was
decided to devote a small percentage of the observing time of this space
telescope with an effective 95-cm aperture to asteroseismology. Kepler is
the most powerful photometry tool for asteroseismology to date and
observes a $\sim 10 \times 10$ degree field for three years,
practically without interruption, providing a time base considerably
longer than all other present missions. Initial results of the Kepler
Asteroseismic Investigation, based on the first 43 days of science
operations, were summarized by Gilliland et al.\ (2010).

A possible shortcoming of all these asteroseismic space missions is that 
they observe in only one passband. Therefore all information available 
for seismic modelling are the targets' oscillation frequencies, unless 
mode identifications are provided by ground-based support observations. 
These are often difficult because the larger space telescopes mostly 
observe faint targets. The BRITE-Constellation (BRIght Target Explorer, 
http://www.brite-constellation.at) mission therefore adopts quite a 
different strategy. This mission consists of three pairs of 
nanosatellites hosting a 3-cm telescope each that will observe in at 
least two different passbands (one per satellite), facilitating mode 
identification with the photometric method. BRITE-Constellation will 
preferentially observe stars brighter than 5th magnitude and has a large 
$24 \times 24$ degree field of view. Given the brightness of the science 
targets, mode identification from high-resolution spectroscopy can also 
easily be done. The first pair of satellites are to be launched in early 
2011.

Finally, PLATO (PLAnetary Transits and Oscillations, 
http://sci.esa.int/plato) is a mission designed to provide a full 
statistical analysis of exoplanetary systems around nearby and bright 
stars. Currently in the definition phase, it will host 28 telescopes of 
10~cm aperture that will observe two 557 square degree fields for 2.5 
years each. It is intended to observe 100000 stars to a precision of 
1~ppm per month and 500000 stars to somewhat poorer accuracy to 
determine stellar and planetary masses to 1\%.

\subsection*{Prospects and problems}

Asteroseismology is a research field evolving so rapidly that some 
of the results reported here will already be outdated when this book 
appears in print. Given the $\sim$~30-year headstart of helioseismology 
in comparison, the field now is in its teenage years, but matures 
rapidly. 

The theoretical basis for asteroseismic studies is laid, although far 
from being perfect. Some of the problems that require solution comprise 
improved treatment of magnetic fields, convection, internal flows, and 
fast rotation. It is still poorly known what causes stellar cycles, and 
what makes certain classes of pulsator select the types of mode they 
oscillate in. Some asteroseismic results point towards a requirement of 
still higher heavy-element opacities, which current calculations do not 
seem capable of providing.

Observationally, pulsating white dwarf stars, $\beta$~Cephei stars and 
solar-like oscillators have been studied asteroseismically, and continue 
to be. There are high hopes that present asteroseismic space missions 
will open the $\delta$~Scuti, SPB, $\gamma$~Doradus and V1093 Herculis 
stars for interior structure modelling, and further improve the 
situation for V361 Hydrae stars. Future high-precision radial velocity 
networks and sites will improve our knowledge mostly for solar-like 
oscillators and roAp stars, with the latter guiding theory of stellar 
pulsation under the influence of rotation and magnetic fields. The 
Kepler mission will provide asteroseismic results for solar-like 
oscillators en masse, and a large number of massive stars are expected 
to be studied with BRITE-Constellation. The Gaia mission 
(http://www.rssd.esa.int/gaia) is expected to provide accurate 
luminosity determinations for a vast number of asteroseismic targets, 
tightly constraining the modelling.

It is therefore only appropriate to finish with a quote by Eyer \& 
Mowlavi (2008): now is the time to be an asteroseismologist!

\section*{Acknowledgements}

I am grateful to Victoria Antoci, Tim Bedding, G\"unter Houdek and Mike 
Montgomery for their comments on this manuscript, as well as to Joris De 
Ridder and Thomas Lebzelter for helpful input. Tim Bedding and Patrick 
Lenz provided some results reproduced here. I apologize to all 
colleagues whose work was not properly cited here due to a rigorous 
restriction on the total number of literature sources to be quoted in 
this article.

\section*{References and resources}

\vspace{2mm} \noindent Aerts, C., 2007, {\it Lecture Notes on Asteroseismology}, 
Katholieke Universiteit Leuven

\vspace{2mm} \noindent Aerts, C., Thoul, A., Daszy{\'n}ska, J., Scuflaire, R., 
Waelkens, C., Dupret, M. A., Niemczura, E., \& Noels A., 2003, Science, 
300, 1926

\vspace{2mm} \noindent Aizenman, M., Smeyers, P., \& Weigert, A., 1977, A\&A 58, 41

\vspace{2mm} \noindent Asplund, M., Grevesse, N., Sauval, A., Allende Prieto, C., \& 
Kiselman, D., 2004, A\&A, 417, 751

\vspace{2mm} \noindent Astronomy \& Astrophysics, 2009, Special Issue, Vol. 506

\vspace{2mm} \noindent Baglin, A. 2003, Advances in Space Research, 31, 345

\vspace{2mm} \noindent Baker, N., \& Kippenhahn, R., 1962, Zeitschrift f\"ur
Astrophysik, 54, 114

%\vspace{2mm} \noindent Basu, S., 2009, Ap\&SS, in press

\vspace{2mm} \noindent Bedding, T. R., et al., 2004, ApJ 614, 380

\vspace{2mm} \noindent Bedding, T. R., et al., 2007, ApJ 663, 1315

\vspace{2mm} \noindent Belkacem, K., et al., 2009, Science, 324, 1540

\vspace{2mm} \noindent Brassard, P., Fontaine, G., 2005, ApJ 622, 572

\vspace{2mm} \noindent Breger, M., Lenz, P., \& Pamyatnykh, A. A., 2009, MNRAS, 396, 291

\vspace{2mm} \noindent Brickhill, A. J., 1991, MNRAS 251, 673

\vspace{2mm} \noindent Bruntt, H., \& Southworth, J., 2008, Journal of Physics: 
Conference Series Vol. 118, 012012

\vspace{2mm} \noindent Campbell, W. W., Wright, W. H., 1900, ApJ, 12, 254

\vspace{2mm} \noindent Chaplin, W. J., Houdek, G., Karoff, C., Elsworth, Y. \& New, 
R., 2009, A\&A, 500, L21

\vspace{2mm} \noindent Charpinet, S., Brassard, P., Fontaine, G., Green, 
E. M., Van 
Grootel, V., Randall, S. K., Chayer, P., 2009, in {\it Stellar Pulsation: 
Challenges for Observation and Theory}, eds. J. A. Guzik \& P. A. 
Bradley, AIP Conference Proceedings, Vol.\ 1170, p.\ 585

\vspace{2mm} \noindent Christensen-Dalsgaard, J., 1988, in {\it Advances in Helio- and 
Asteroseismology}, ed. Christensen-Dalsgaard, J., \& Frandsen, S., Proc.\ 
IAU Symp.\ 123, p.\ 295

\vspace{2mm} \noindent Christensen-Dalsgaard, J., 2002, Rev.\ Mod.\ Phys.\ 74, 1073

\vspace{2mm} \noindent Christensen-Dalsgaard, J., 2010, {\it Lecture Notes on Stellar 
Oscillations}, \AA rhus University

\vspace{2mm} \noindent Cox, J. P, 1984, PASP 96, 577

\vspace{2mm} \noindent Daszy{\`n}ska-Daszkiewicz, J., Dziembowski, W. A.,
Pamyatnykh, A. A., \& Goupil, M.-J., 2002, A\&A, 392, 151

\vspace{2mm} \noindent Daszy{\`n}ska-Daszkiewicz, J., Dziembowski, W. A., \& 
Pamyatnykh, A. A., 2005, A\&A, 441, 641

\vspace{2mm} \noindent Daszy{\`n}ska-Daszkiewicz, J., Walczak, P., 2009, MNRAS, 398, 
1961

\vspace{2mm} \noindent De Ridder, J., et al., 2004, MNRAS, 351, 324

\vspace{2mm} \noindent Deubner, F.-L., \& Gough, D. O., 1984, Ann.\ Rev.\ Astron.\ 
Astroph., 22, 593

%\vspace{2mm} \noindent Deutsch, A. J., 1958, in {\it Electromagnetic 
%Phenomena in Cosmical Physics}, ed. B. Lehnert, Proc.\ from IAU Symp. 
%6, p.\ 209

\vspace{2mm} \noindent Dufour, P., Green, E. M., Fontaine, G., Brassard, P., Francoeur, 
M., Latour, M., 2009, ApJ, 703, 240

\vspace{2mm} \noindent Dupret, M.-A., Grigahc{\`e}ne, A., Garrido, R., Gabriel, M., 
\& Scuflaire, R., 2005, A\&A, 435, 927

\vspace{2mm} \noindent Dziembowski, W. A., 1977, Acta Astron., 27, 203

\vspace{2mm} \noindent Dziembowski, W. A., \& Pamyatnykh, A. A., 1991, A\&A, 248, L11

\vspace{2mm} \noindent Dziembowski, W. A., \& Goode, P. R., 1992, ApJ, 394, 670

\vspace{2mm} \noindent Dziembowski, W. A., \& Pamyatnykh, A. A., 2008, MNRAS, 385, 2061

\vspace{2mm} \noindent Eyer, L., Mowlavi, M., 2008, J.\ Phys.\ Conf.\ Ser.\ 118, 012010

\vspace{2mm} \noindent Fabricius, D., 1596, reported by Wolf, R. 1877, 
{\it Geschichte der Astronomie}, Verlag R. Oldenbourg, Munich, p.\ 116

\vspace{2mm} \noindent Fleming, W. P., 1899, reported by Pickering, E. 
C., et al., 1901, ApJ, 13, 226

\vspace{2mm} \noindent Frandsen, S., et al., 2002, A\&A, 394, L5

\vspace{2mm} \noindent Frost, E. B., 1902, ApJ, 15, 340

\vspace{2mm} \noindent Gautschy, A., Saio, H., 1995, Ann.\ Rev.\ Astron.\ Astroph.\ 33, 
75

\vspace{2mm} \noindent Gilliland, R. L., et al., 2010, PASP, 122, 131

\vspace{2mm} \noindent Gizon, L., Birch, A. C., \& Spruit, H. C., 2010, Ann.\ Rev.\ 
Astron.\ Astroph., submitted (arXiv:1001.0930)

\vspace{2mm} \noindent Goodricke, J., 1786, Phil.\ Trans.\ R.\ Soc.\ 
Lond., 76, 48

\vspace{2mm} \noindent Gough, D. O., 1986, in {\it Hydrodynamic and Magnetodynamic 
Problems in the Sun and Stars}, ed.\ Y. Osaki, Tokyo University, p.\ 117

\vspace{2mm} \noindent Gough, D. O., 1996, The Observatory, 116, 313

\vspace{2mm} \noindent Green, E. M., et al., 2003, ApJ, 583, L31

\vspace{2mm} \noindent Guzik, J.\,A., Keady, J.\,J., \& Kilcrease, D.\,P., 2009, in {\it 
Stellar Pulsation: Challenges for Observation and Theory}, ed.\ J.\,A.\ 
Guzik \& P.\,A.\ Bradley, AIP Conf.\,Proc.\,Vol.\,1170, p.\,577

\vspace{2mm} \noindent Handler, G., 1999, MNRAS, 309, L19

\vspace{2mm} \noindent Handler, G., et al., 2002, MNRAS, 333, 262

\vspace{2mm} \noindent Handler, G., et al., 2006, MNRAS, 365, 327

\vspace{2mm} \noindent Handler, G., et al., 2009, ApJ, 698, L56

\vspace{2mm} \noindent Hekker, S., et al., 2009, A\&A, 506, 465

\vspace{2mm} \noindent Herschel, W., 1782, reported by Pickering, E. C., 
1884, Proc.\ of the American Academy of Arts and Sciences, Vol.\ 19, p.\ 269

\vspace{2mm} \noindent Houdek, G., in {\it Stellar Pulsation: Challenges for 
Observation and Theory}, eds. J. A. Guzik \& P. A. Bradley, AIP 
Conference Proceedings, Volume 1170, p.\ 519

\vspace{2mm} \noindent Jeffery, C. S., 2008, Inf.\ Bull.\ Var.\ Stars, No.\ 5817

\vspace{2mm} \noindent Jones, P. W., Hansen, C. J., Pesnell, W. D., \& Kawaler, S. 
D., 1989, ApJ, 336, 403

\vspace{2mm} \noindent Kallinger, T., \& Matthews, J. M., 2010, ApJ, 711, L35

\vspace{2mm} \noindent Kaye, A. B., Handler, G., Krisciunas, K., 
Poretti, E., \& Zerbi, F. M., 1999, PASP, 111, 840

\vspace{2mm} \noindent Kepler, S. O., et al., 2005, ApJ, 634, 1311

\vspace{2mm} \noindent Kilkenny, D., Koen, C., O'Donoghue, D., \& 
Stobie, R. S., 1997, MNRAS, 285, 640

\vspace{2mm} \noindent Kim, A., Montgomery, M. H., \& Winget, D. E., 2006, in {\it New 
Horizons in Astronomy}, ed.\ S. J. Kannappan et al., ASP Conf.\ Ser., 352, 
253

\vspace{2mm} \noindent Kiss, L. L., Szab{`o}, G. M., \& Bedding, T. R., 2006, MNRAS 
372, 1721

\vspace{2mm} \noindent Kjeldsen, H., Bedding, T. R., Viskum, M., \& Frandsen, S., 
1995, AJ 109, 1313

\vspace{2mm} \noindent Kurtz, D. W., 1982, MNRAS 200, 807

\vspace{2mm} \noindent Kurtz, D. W., 2000, in {\it Delta Scuti and Related Stars}, ed.  
M. Breger \& M. H. Montgomery, ASP Conf. Ser. Vol. 210, p. 287

\vspace{2mm} \noindent Kurtz, D. W., 2009, in {\it Stellar Pulsation: 
Challenges for Observation and Theory}, eds. J. A. Guzik \& P. A. 
Bradley, AIP Conference Proceedings, Volume 1170, p.\ 491

\vspace{2mm} \noindent Kurtz, D. W., \& Martinez, P., 2000, Baltic Astronomy, 9, 253

\vspace{2mm} \noindent Kurtz, D. W., Shibahashi, H., Dhillon, V. S., Marsh, T. R., 
Littlefair, S. P., 2008, MNRAS, 389, 1771

\vspace{2mm} \noindent Landolt, A. U., 1968, ApJ, 153, 151

%\vspace{2mm} \noindent Lenz, P., Pamyatnykh, A. A., Breger, M., 2007, 
% in {\it Unsolved 
% Problems in Stellar Physics}, AIP Conf.\ Proc., Vol.\ 948, p.\ 201

\vspace{2mm} \noindent Maeder, G., \& Meynet, G., 2000, Ann.\ Rev.\ Astron.\ 
Astroph., 38, 143

\vspace{2mm} \noindent Matthews, J. M., 2006, in {\it Highlights of Recent Progress 
in the Seismology of the Sun and Sun-Like Stars}, 26th meeting of the 
IAU, Joint Discussion 17, \#21

\vspace{2mm} \noindent McGraw, J. T., Liebert, J., Starrfield, S. G., \& 
Green, R., 1979, in {\it White dwarfs and variable degenerate stars}, 
University of Rochester, p.\ 377

\vspace{2mm} \noindent Metcalfe, T. S., 2005, MNRAS, 363, L86

\vspace{2mm} \noindent Metcalfe, T. S., Montgomery, M. H., \& Kanaan, A. 2004, ApJ 605, 
L133

\vspace{2mm} \noindent Miglio, A., Montalb{\'a}n, J., Noels, A., \& Eggenberger, P., 
2008, MNRAS, 386, 1487

\vspace{2mm} \noindent Mkrtichian, D. E., Hatzes, A. P., Saio, H., \& Shobbrook, R. R., 
2008, A\&A, 490, 1109

\vspace{2mm} \noindent Montagner, J.-P., Roult, G., 2008, Journal of 
Physics: Conference Series Vol. 118, 012004

\vspace{2mm} \noindent Montgomery, M. H., 2005, ApJ, 633, 1142

\vspace{2mm} \noindent Montgomery, M. H., 2009, in {\it Stellar Pulsation: Challenges 
for Observation and Theory}, eds. J. A. Guzik \& P. A. Bradley, AIP 
Conference Proceedings, Volume 1170, p. 605

\vspace{2mm} \noindent Montgomery, M. H., Williams, K. A., Winget, D. E., Dufour, P., 
De Gennaro, S., Liebert, J., 2007, ApJ, 678, L51

\vspace{2mm} \noindent Mosser, B., et al., 2009, Comm.\ Asteroseis., 158, 337

\vspace{2mm} \noindent Mukadam, A., et al., 2007, ApJ, 667, 433

\vspace{2mm} \noindent Nather, R. E., \& Robinson, E. L., 1974, ApJ, 190, 637

\vspace{2mm} \noindent Nather, R. E., Winget, D. E., Clemens, J. C., Hansen, C. J., \&
Hine, B. P., 1990, ApJ, 361, 309

\vspace{2mm} \noindent Pamyatnykh, A. A., Handler, G., \& Dziembowski, W. A., 2004, 
MNRAS 350, 1022

\vspace{2mm} \noindent Percy, J. R., {\it Understanding Variable Stars}, 2007, 
Cambridge University Press

\vspace{2mm} \noindent Pesnell, W. D., 1985, ApJ 292, 238

\vspace{2mm} \noindent Reed, M. D., Brondel, B. J., \& Kawaler, S. D., 2005, ApJ 634, 
602

\vspace{2mm} \noindent Reese, D. R., Thompson, M. J., MacGregor, K. B. Jackson, S.,
Skumanich, A., \& Metcalfe, T. S., 2009, A\&A, 506, 183

\vspace{2mm} \noindent Robinson, E. L., Kepler, S. O., \& Nather, R. E., 1982, ApJ, 
259, 219

\vspace{2mm} \noindent Rosseland, S., \& Randers, G., 1938, Astrophisica Norvegica, 3, 71

\vspace{2mm} \noindent Saio, H., et al., 2006, ApJ, 650, 1111

\vspace{2mm} \noindent Samadi, R., Goupil, M.-J., \& Houdek, G., 2002, A\&A, 395, 563

%\vspace{2mm} \noindent Stankov, A., \& Handler, G., 2005, ApJS, 158, 193

\vspace{2mm} \noindent Suran, M., Goupil, M., Baglin, A., Lebreton, Y., \& Catala, 
C., 2001, A\&A, 372, 233

\vspace{2mm} \noindent Talon, S., Zahn, J.-P., Maeder, A., \& Meynet, G., 1997, A\&A, 
322, 209

\vspace{2mm} \noindent Tassoul, M., 1980, ApJS, 43, 469

\vspace{2mm} \noindent Telting, J. H., 2003, Ap\&SS, 284, 85

\vspace{2mm} \noindent Telting, J. H., \& Schrijvers, C., 1997, A\&A 317, 723

\vspace{2mm} \noindent Torres, G., Andersen J., \& Gim{\`e}nez, A., 2010, A\&ARv 18, 
67

\vspace{2mm} \noindent Townsend, R., 2009, in {\it Stellar Pulsation: Challenges 
for Observation and Theory}, eds. J. A. Guzik \& P. A. Bradley, AIP 
Conference Proceedings, Volume 1170, p. 355

\vspace{2mm} \noindent Unno W., Osaki Y., Ando H., Saio H., Shibahashi H., 1989, 
{\it Nonradial Oscillations of Stars}, University of Tokyo press, 2nd 
edition

\vspace{2mm} \noindent Vogt, S. S., Penrod, G. D., 1983, PASP, 95, 565

\vspace{2mm} \noindent Waelkens, C., Aerts, C., Kestens, E., Grenon, M., Eyer, L., 
1998, A\&A 330, 215

\vspace{2mm} \noindent Waelkens, C., \& Rufener, F., 1985, A\&A 152, 6

\vspace{2mm} \noindent Walker, G., et al., 2003, PASP, 115, 1023

\vspace{2mm} \noindent Watson, R. D., 1988, Ap\&SS 140, 255

\vspace{2mm} \noindent Winget, D. E., Robinson, E. L., Nather, R. E., \& Fontaine, G., 
1982, ApJ, 262, L11

\vspace{2mm} \noindent Winget, D. E., et al., 1991, ApJ, 378, 326

\vspace{2mm} \noindent Winget, D. E., et al., 1994, ApJ, 430, 839

\vspace{2mm} \noindent Winget, D. E., Sullivan, D. J., Metcalfe, T. S., Kawaler, S. 
D., \& Montgomery, M. H., 2004, ApJ, 602, L109

\vspace{2mm} \noindent Woudt, P. A., et al., 2006, MNRAS, 371, 1497

\end{document}